\documentstyle[11pt,aaspp4,psfig]{article}

\lefthead{Chiappini et al.}
\righthead{Chemical Evolution of the Galaxy with Variable IMFs} 

\begin{document}

\title{The Chemical Evolution of the Galaxy with Variable IMFs}

\author{C. Chiappini\altaffilmark{1}, F. Matteucci\altaffilmark{2}}

\affil{Departamento de Astronomia, Observat\'orio Nacional, R. Gal. Jos\'e Cristino, 77 - Rio de Janeiro - RJ, CEP 20921-400, Brasil}
\affil{Dipartimento di Astronomia, Universit\'a di Trieste,
SISSA, Via Beirut 2-4, I-34013 Trieste, Italy}

\and

\author{P. Padoan\altaffilmark{3}}
\affil{Department of Astronomy, Harvard University, Cambridge, MA 02138, USA}





\begin{abstract}
In this work we explore the effects of adopting initial mass functions
(IMFs) variable in time on the chemical evolution of the Galaxy.
In order to do that we adopt a chemical evolution model which assumes
two main infall episodes for the formation of the Galaxy and which proved
to be successful in reproducing the majority of the observational constraints, 
at least for the case of a constant IMF.

Different variable IMFs are tested with this model, all assuming that massive stars
are preferentially formed in ambients of low metallicity. This implies that massive stars 
are formed preferentially at early times and at large galactocentric distances.

Our numerical results have shown that all the variable IMFs proposed so far are 
unable to reproduce all the relevant observational constraints in the Galaxy
and that a constant IMF still reproduces better the observations. In particular,
variable IMFs of the kind explored here are unable to reproduce the observed
abundance gradients even when allowing for changes in other chemical evolution model
parameters as, for instance, the star formation rate.

As a consequence of this we conclude that the G-dwarf metallicity distribution is
best explained by infall with a large timescale and a constant IMF, since it is
possible to find variable IMFs of the kind studied here, reproducing the G-dwarf metallicity but this 
worsen the agreement with other observational constraints.

\end{abstract}

\section{Introduction}

The initial mass function (IMF) is one of the basic ingredients of 
chemical evolution models. 
The IMF is normally derived from the observed present day mass function 
(PDMF) in the solar neighborhood and is assumed to be independent of time. 
The PDMF gives the {\it present} number of main sequence (MS) stars with mass
between $m$ and $m+dm$ per $pc^2$ per $\log m$ and is obtained from the
observed luminosity function of MS stars (see Scalo 1986).
 
The derivation of the IMF from the PDMF is a difficult procedure involving
assumptions about the behaviour of the star formation rate during the
lifetime of the Galaxy (see Tinsley 1980; Scalo 1986). In particular,
for stars with lifetimes larger than the age of the Galaxy 
($m \leq 1 M_{\odot}$) the IMF is derived by assuming an average 
star formation rate in the past,
whereas for stars with lifetimes negligible relative to the age of the Galaxy
($m \geq 2 M_{\odot}$) the IMF is derived by a assuming a present time star
formation rate and taking into account the stellar lifetimes ($\tau_m$).
The IMF in the mass range between 1 and 2 $M_{\odot}$ is obtained by connecting
the two parts of the IMF described above. It is worth noting that in all 
this procedure the main assumption is that the IMF has not varied in time.
The IMF is usually approximated by a power law of the form 
$\phi(m) \propto m^{-(1+x)}$ with $m_{low} \leq m \leq m_{upper}$.

This method has been often adopted and, since it cannot enable a precise
determination of the IMF slope,
different results have been reported.
Salpeter (1955), in a pioneering work, found a slope $x=1.35$ in 
the mass range 2-10 $M_{\odot}$. For the sake of simplicity this IMF has been 
often extrapolated to upper and lower
masses in order to be used in chemical evolution models.
More complex power laws, with different slopes for different mass ranges,
have been proposed (Miller \& Scalo 1979; Tinsley 1980; Scalo 1986 and
Kroupa et al. 1993). All the IMFs discussed above are constant in time.

However, a variable IMF, which formed relatively more massive stars during the
earlier phases of the evolution of the Galaxy compared to the one
observed today in the solar vicinity, has often been suggested as being 
one of the possible solutions for the G-dwarf problem (namely the deficiency 
of metal-poor stars in the solar neighborhood
when compared with the number of such stars predicted by the simple model).
Such an IMF would also be physically plausible from the theoretical point
of view 
if the IMF depends on a mass scale such as the Jeans mass. 
In fact, the Jeans mass is strongly dependent on the temperature which would have been higher
at earlier times (see Larson 1998).
Terlevich \& Melnick (1983) had proposed an 
IMF in which the slope $x$ is a strong function of the metallicity. However
the parameterization for this dependence was very uncertain.
This IMF was used in a chemical evolution model by Matteucci \& Tornamb\'e 
(1985) who suggested that only an IMF in which the slope is a less
strong function of the metallicity could lead to an agreement with
the observed properties of our Galaxy. More recent suggestions about 
metallicity dependent IMFs are due to Chuvenkov \& Gluknov (1995),
Silk (1995) and Scully et al.(1996). 

Ferrini et al. (1990) investigated the problem of the origin of the IMF by
assuming a non-linear and non-equilibrium dynamics when describing the 
fragmentation of molecular clouds.
In such a system, the star formation occurs if an instability criterium is 
satisfied giving rise to the mass spectrum of the newly formed stars. 
Taking into account different instability criteria when defining the critical
mass for fragmentation, these
authors have shown that in order to explain the IMF for the solar 
vicinity one has to assume the more realistic view that the critical mass for
fragmentation is determined by a cooperative effect of different forces,
namely, gravitational, magnetic and turbulent.
These calculations however
cannot give any information about the time dependence of the IMF but only
produce an IMF in agreement to the one of the solar neighborhood. 

Another theoretical approach to the IMF problem is the one proposed
by Padoan et al. (1997 - hereinafter PNJ). Since random motions are probably ubiquitous 
in sites of star formation, PNJ suggested
 to describe the formation of protostars as the 
gravitational collapse of Jeans masses in a density distribution 
shaped by random supersonic motions. Such flows create a highly nonlinear 
density field, with density contrasts of a few orders of magnitudes. 
They show that the  probability density function ({\it pdf}) of the density 
field can be well approximated by a Log-Normal distribution.
The weakness of such an IMF is the fact that it predicts that the most massive
stars should form where the density is smallest, contrary to what is observed
in star-forming regions (see Scalo et al. 1998 for a detailed discussion on this point and further critics also on theoretical grounds).

On the empirical grounds there is at present no clear direct evidence that the IMF 
in the Galaxy has varied with time.
A detailed discussion about possible observed variations
in the IMF in different environments can be found in the recent review
by Scalo (1998), but such variations are comparable with the uncertainties
still involved in the IMF determinations. 
The present uncertainties in the observational results
prevent any conclusion against the idea of an universal IMF.

Given the uncertainties in both theoretical and observational grounds,
the IMF variations can be treated in a parameterized form and the
proposed IMFs can be tested by means of a detailed chemical evolution model.
We chose to rely on the so-called two-infall model
(Chiappini et al. 1997 - hereinafter CMG97) that proved to provide results in good agreement with 
several observational constraints in the Milky Way.
The observational constraints
are of fundamental importance to build a realistic chemical evolution model and
this is why the Milky Way is a good starting point because for our galaxy we
have a high number of observational constraints available.
With respect to these constraints the last years have been of crucial importance and,
in the case of the Milky Way, the new observational data required a revision
of the previous chemical evolution models (CMG97,
Pagel \& Tautvaisiene 1995, Pagel 1997). The results obtained by CMG97 from a
careful comparison between model predictions and observational constraints strongly
suggest that the previously adopted picture for the Galaxy formation in which
the gas shed from the halo was the main contributor to the thin disk formation,
is not valid anymore.

However, the two-infall model adopted a constant
IMF. The aim of this paper is to address the question of what time-dependent 
IMF properties are allowed in order to still match the observational
constraints when adopting the two-infall model (CMG97).
To do this we test different IMFs in our chemical evolution code, going
from those where the variability is contained on the 
slope of the power-law which is taken as a function
of the metallicity (Scully et al. 1996, Terlevich \& Melnick 1983) to the 
one by PNJ which predicts also a change in the stellar mass which
contributes most to the IMF as a function of time.

In section 2 we discuss the characteristics of the IMFs adopted. In the 
particular case of the PNJ IMF we show the assumptions necessary to
apply this function to the Galactic disk. 
In section 3 we describe briefly the chemical evolution model adopted. In section 4 we present
the results and the conclusions are drawn in section 5.

\section{The Variable IMFs adopted}

\subsection{PNJ IMF}

In the approach adopted by PNJ
the IMF contains a dependence 
on the average physical parameters 
(temperature, density, velocity dispersion of gas) of large
scale sites of star formation. 
The model is based on numerical experiments
of highly supersonic random flows (Nordlund \& Padoan 1997).
In this model the most probable stellar mass per logarithmic mass interval, that is
the stellar mass which contributes most to the mass function, defined as
$x(M_{max}) \equiv 0$, is given by:

\begin{eqnarray}
M_{max}&\!\!\!\!=&\!\!\!\! 
0.2M_{\odot} \!\!\left(\frac{n}{1000cm^{-3}}\right)^{-1/2}
 \!\!\left(\frac{T}{10 K}\right)^{2} \!\!\left(\frac{\sigma_{v}}{2.5 km/s}\right)^{-1}
\nonumber\\
& &\label{17} 
\end{eqnarray}

As we will see in the next sections, this IMF favors higher mass stars in the 
earlier times when applied to the chemical evolution of the Milky Way.
We would like to point out that one of the possible problems of the equation quoted 
above is the fact that
such an IMF predicts that the most massive stars form in the lowest-density
and hottest regions which seems to violate some theoretical expectations.
For a careful discussion on this particular point see Scalo et al. (1998). These
authors call attention to the fact that at present there is still no physical basis
for such an IMF and that other IMFs besides the one adopted here could be constructed
to give similar dependences. 

As it is clear from the equation (1), in order to apply the PNJ IMF, 
the sites of star formation must be described in terms of the values of 
the velocity dispersion, temperature and mean density. 
As the chemical evolution model adopted here does not contain
the energetics of the ISM, some simplifying hypothesis are needed in order
to apply this IMF to the two-infall model.
In order to do so we assume that
most of the stars are born in giant molecular cloud complexes, namely in cold 
gas at temperature $T=10K$, mean density $<n>=50-100 cm^{-3}$, and typical
size of 100 pc (Dame et al. 1986). 
This is almost true at the present time (Myers et al. 1986, Mooney \& Solomon 1988, Gerritsen \& Icke 1997), 
but could be a weak hypothesis during the early phases of the formation
of the disk.

In fact, the adopted mean cloud temperature could be uncertain even for the present as
it holds only for clouds with virtually no star formation. Moreover, as the temperature of molecular 
clouds is controlled by the balance between heating by external radiation fields
and cooling by line and dust emission we should expect that it
varied with time. In principle, the effect of changes
in temperature can be compensated by sufficiently large changes in pressure or
gas density (see Larson 1998). However, as discussed by Larson it is much easier
to find reasons to expect higher molecular cloud temperatures  at earlier times than
important pressure or density variations.  

As it will be discussed in the next sections our hypothesis of constant
molecular temperature and density throughout the evolution of the Galaxy could be 
the responsible for the mild variation of the PNJ IMF
when applied to the two-infall model. 
In fact,
the main effect of such hypothesis will be to make the IMF variable only in the
very early phases of the disk evolution. 

Once these values of temperature and mean density are adopted, the IMF depends
on the particular value of the velocity dispersion of the large scale site of star
formation. As an estimate of the velocity dispersion of the cold gas, we use the 
vertical velocity dispersion (perpendicular to the plane of the galactic disk), under
the assumption that the gaseous disk (cold molecular gas) is supported in the vertical
direction by the velocity dispersion of molecular clouds.
In fact, the vertical velocity dispersion is expected to be comparable with the
internal velocity dispersion of giant molecular cloud complexes because, 
according to Larson (1981),  the velocity
dispersion of gas generally scales as the squared root of 
the linear dimension of the considered region and the giant molecular cloud complexes have 
a linear extension comparable with the scale height of the disk.

Following Bahcall \& Casertano (1984), we can relate the total surface 
mass density
$\sigma_{tot}$, the scale-height $h$, and the vertical velocity dispersion $\sigma_{v}$:

\begin{equation}
\sigma_{v}^2\approx 2\pi G h \sigma_{tot}
\end{equation}
This is a good approximation in the early evolution of the disk, when the 
gravitational potential is dominated by the gas, while it is a rough approximation
towards the end of the disk evolution, when the stellar component contributes
significantly to the gravitational potential and its scale height can grow
larger than the scale height of the cold gas. N-body simulations by
Gerritsen \& Icke (1997) show that the star formation can only occur in the mid plane
of a spiral galaxy where the gas is dense enough to cool, and that for this gas the 
scale height is lower than the one of the other components, namely old stars or warm gas.

To be able to predict the vertical velocity dispersion
of the clouds, we assume that: {\it i)} the disk surface density has a radial distribution 
and a time evolution determined by an infall law. The radial
distribution is exponential (see CMG97) and that {\it ii)} the scale height of the molecular clouds in the disk
is constant with time and radius, $h=100 pc$.

The second of the two hypothesis above is based on a
mixture of theoretical assumptions and observational 
evidences. We assume $h$ to be constant in time
because its time evolution during the disk formation process
is poorly known. However, the vertical velocity dispersion grows 
only as the square root of $h$, so even if the scale height
were initially 4 times larger, the velocity dispersion would
be only twice as large as assumed here. We have verified
that the results of the present work are not significantly 
affected by such a time variation of the velocity dispersion.
We also assume $h$ to be constant along the radius because
a number of observational results in our Galaxy and in external
edge-on late type galaxies have shown that the scale
height is approximately constant in the molecular gas (Solomon, 
Sanders \& Scoville 1979; Sanders, Solomon \& Scoville 1984; 
Garc\'{\i}a-Burillo, Dahlem \& Gu\'{e}lin 1991; Heiles 1991), 
in the atomic gas (Dickey \& Lockman 1990; Heiles 1991; 
Ferrara 1993, 1996), and in the stellar disk 
(de Grijs \& Peletier 1997). By approximately 
constant we mean that the variation of the scale height is small
compared with the exponential radial variation of the column 
density. However, 
all of these considerations refer mostly to the region inside 10 kpc
from the galactic center, since in the
outer disk both the molecular and the atomic gas components flare
to large scale height (Grabelsky et al. 1987). 
Therefore, our assumption weakens outside the region of 10 Kpc.

Once $h$ is assumed to be constant, the velocity dispersion of the 
molecular clouds grows towards the center of the disk, according to
equation (2). 

Since the total surface mass density of the disk is assumed to have an 
exponential dependence on the galactocentric distance and to increase 
with time (see the following section), the vertical velocity dispersion 
depends on galactocentric distance, $r$, and time $t$:
\begin{equation}
\sigma_{v}(r,t)\approx 3.8km/s\left( h_{kpc}\sigma_{tot}(r,t)\right)^{1/2}
\end{equation} 
\noindent
where $h_{kpc}$ is the scale-height of the (cold gas) disk in $kpc$
and
$\sigma_{tot}(r,t)$ is the total (gas plus stars) surface mass
density of the disk, measured in $M_{\odot}/pc^2$.

Under these assumptions, the IMF in the galactic disk can be determined. 

From eq. (1), the typical stellar mass is inversely proportional to the
velocity dispersion of the star forming gas. Since the surface density
has an exponential distribution (at least approximately in the present model),
the velocity dispersion also depends exponentially on the galactocentric distance, 
with a scale-length equal to twice the value for the disk mass density, and
the typical stellar mass grows exponentially with galactocentric distance,
with the same scale-length as the velocity dispersion.

Given the mild variation of the PNJ IMF when applied to the two-infall model, 
due to our assumptions of constant molecular temperature
and density, we will explore cases of more strongly variable
IMFs in a parameterized form in the next sections.
In fact, in order to follow the evolution of the gas temperature we would need
an hydrodynamical model which is beyond the scope of this paper.

\subsection{IMFs with a variable slope}

Following a suggestion by Terlevich \& Melnick (1983), Matteucci
and Tornamb\'e adopted a parameterized variable IMF assuming its slope $x$ to be
a function of the metal content of the interestellar medium:
$x = logZ + 0.45$
for $ Z > 0.002$ whereas for $Z \leq 0.001 $ a Salpeter IMF was adopted.
This IMF was tested in a model which formed the disk in a much smaller
timescale than the one adopted in the present two-infall model.

The third IMF we will adopt is the one proposed by Scully et al. (1986).
This IMF also has a slope which is a function of metallicity, 
parameterized as a function oxygen
instead of the global total metal content. The following parameterization is adopted:
$x = 1.25 + O/O_{\odot}$.

In previous works, both these IMFs where tested only in the solar vicinity. In the next sections we will
see also the predicted gradients for the abundance of oxygen and the gas in the 
disk of the Milky Way when such IMFs are applied to the two-infall model.

\section{The Model}

The two infall model assumes two major
gas infall episodes. The first one is responsible for the formation of the
halo and part of the thick disk. The second infall episode, delayed
with respect to the first, forms the thin disk. One of the basic results
of CMG97 is that in order to reproduce simultaneously 
all the constraints
in our Galaxy one should disentangle the evolution of the halo-thick
disk from that of the thin disk. This implies that most of the thin disk
was formed by accretion of extragalactic primordial material 
and not only from gas lost
from the halo as previously thought. The timescales suggested for the 
formation of the halo/thick disk and the thin disk are 1 Gyr and 8 Gyr, 
respectively. 

As in Matteucci \& Fran\c cois (1989) the Galactic disk is approximated
by several independent rings, 2 kpc wide, without exchange of matter 
between them. We chose not to include radial flows as it is known that
they can not originate an abundance gradient by themselves (Edmunds \& Greenhow 1995). 
In our model the abundance gradient arises, for a constant IMF, as a consequence
of the assumed inside-out formation of the Galactic disk.
The basic equations are the same as in Matteucci \& Fran\c cois (1989), except
for the expression adopted for the rate of mass accretion in each shell which is 
given by:

\begin{eqnarray}
{d \; G_i(r,t)_{\inf} \over dt} &\!\!\!\!=&\!\!\!\! 
 {A(r) \over \sigma (r,t_G)} \; 
(X_i)_{\inf} \;  
e^{-t/\tau_H} \\
&\!\!\!\!+&\!\!\!\! {B(r) \over \sigma (r,t_G)} \; (X_i)_{\inf} \;  
e^{-(t-t_{\max})/\tau_D} \nonumber
\end{eqnarray}
\noindent
where
$G_i(r,t)_{\inf}$ is the normalized surface gas density 
of the infalling material in the form of the element $i$,
${(X_i)}_{\inf}$ gives the composition of the infalling gas, which we 
assume to be primordial,
$t_{\max}$ is the time of maximum gas accretion onto the disk,
and $\tau_{H}$ and $\tau_{D}$ are the timescales for the mass accretion in 
the thick disk and thin disk components, respectively.
These are the two really free parameters
of our model and are constrained mainly by comparison with the observed
metallicity distribution in the solar vicinity. The $t_{\max}$ value
is chosen to be 2 Gyrs and roughly corresponds to the end of the thick
disk phase.
The quantities $A(r)$ and $B(r)$
are derived by the condition of reproducing the current total 
surface mass density distribution in the solar neighbourhood.
The current total surface mass distribution is taken from Rana (1991), whereas
the surface gas density distribution is predicted by the model. 
For the thin disk we assume
a radially varying $\tau_{D}(r)$ which implies that the inner parts of the
thin disk are built much more rapidly than the outer ones. In other words, 
we are dealing with an 
inside-out picture, as suggested by
dynamical models (Larson 1976). The adopted radial dependence of $\tau_{D}$
is:

\begin{equation}
\tau_{D}(r)=0.875r-0.75
\end{equation}

\noindent
As discussed in CMG97 the above expression was built
in order to obtain a timescale for the bulge formation (R $<$ 2kpc) of 1 Gyr,
in agreement with the results of Matteucci \& Brocato (1990) and a 
timescale of 8 Gyr at the solar neighbourhood, which best reproduces
the G-dwarf metallicity distribution (we are adopting $R_{g,\odot}=10 kpc$).

The star formation rate (SFR) adopted here is the same
as in CMG97 and it was chosen in order to give the 
best agreement with the observed constraints. It is assumed to be proportional
both to the surface gas density and total 
surface mass density, with an exponent $k_{HALO}=k_{DISK}=1.5$ 
and a SFR efficiencies of $\nu_{HALO}= 2.0 Gyr^{-1}$ 
and $\nu_{DISK} = 1.0 Gyr^{-1}$ (see CMG97 for details).
Note that the surface mass density exponent of 1.5 obtained from the best
model of CMG97 is in very good agreement with the recent
observational results of Kennicutt (1998). An exponent of this order is also
suggested by the N-body simulations of Gerritsen \& Icke (1997). 

Our adopted SFR has the following form:

\begin{equation}
\Psi(r,t) \propto \nu {\sigma_g}^k {\sigma_{tot}}^{k-1}
\end{equation} 

\noindent
were $\sigma_g$ and $\sigma_{tot}$ are the gas and total surface mass
density respectively.
We also adopted a threshold in the star formation process in which the 
star formation stops when the surface gas density drops
below $ 7 M_{\odot} pc^{-2} $. Such a threshold has been suggested by
star formation studies (Kennicutt 1989, McGaugh \& Blok 1996) 
and naturally produces a gap
in the star formation between the first and second infall episode,
as suggested by observational results discussed in Gratton et al. (1997)
(see CMG97 for a discussion). 

The threshold surface gas density of 7 $M_{\odot}pc^{-2}$, used in the model is not
directly related to the mean density of molecular clouds. Nevertheless, it is
interesting to notice that the existence of this threshold for star formation
and our assumption that stars are formed in molecular clouds are strongly related
to each other. In fact, 7 $M_{\odot}pc^{-2}$ is approximately the column density
necessary for the gas to self-shield from UV radiation and form molecules:
if the disk column density is significantly below this value, molecular clouds
can hardly be formed, and no star formation will occur (if stars are assumed to
be formed in molecular clouds).

The adopted nucleosynthesis 
prescriptions are: i) for low and intermediate mass stars the yields from 
Renzini \& Voli (1981) with $\alpha =1.5$, $\eta =0.33$;
ii) for type Ia SNe the yields from Thielemann et al. (1993); iii) for massive
stars yields from Woosley \& Weaver (1995) and iv) $^{3}$He yields from
Dearborn et al. (1996)

The basic input parameters ($ \nu_H, \nu_D, k_H, k_D, \tau_H, \tau_D $ and
the adopted IMF) are summarized in Table 1. The subscripts H and D refer
to the halo and thin disk, respectively.
 
\section{Results}

The two-infall chemical evolution model, published by
CMG97 (hereinafter Model A) has been modified to 
test the IMF proposed by PNJ (hereinafter Model B and C)
and the other two IMFs (Models D and E). All those models are essentially
the same as model A but with different IMFs.

\subsection{Solar Vicinity}

The PNJ IMF 
predicts a higher number of massive stars in the early phases of the
Galaxy evolution with a progressive increase of low mass stars in 
time. Figure~\ref{fig1a} shows
the behavior of the PNJ IMF at different epochs of the Galactic
evolution. 
From this figure we can see that at the beginning such an
IMF presents a peak at a mass of almost 1 $M_{\odot}$ and 
this peak mass shifts very soon (most of the variation occurs in
the time interval 0.02 - 0.06 Gyr)
towards lower values leading to an IMF of the kind of the one observed
in the solar vicinity.

Figure~\ref{fig1b} and \ref{fig1c} shows the behaviour of the Matteucci \& Tornamb\'e (hereinafter MT) and the Scully et al.
(hereinafter S)
IMFs respectively (models D and E). As can be seen those IMFs vary during all the galactic
lifetime as they are functions of the interstellar metal content. Both IMFs predict a
higher number of massive stars at early times. Only in the very beginning the MT
IMF predicts a lower number of massive stars because of the assumption that 
the IMF is a Salpeter one until the metal abundance in the interstellar medium has reached
a value of 0.002.

Figure~\ref{fig2} shows the effect of the adopted mean density
of the star forming gas on the G-dwarf metallicity 
distribution in model B.
It can be seen that a rather good fit of the G-dwarf metallicity
distribution is obtained in model B, especially when the value $n=50 cm^{-3}$
is adopted for the mean density of the giant molecular clouds. 

Figures~\ref{fig3} and \ref{fig4} compare the model B (with $<n>$=50$cm^{-3}$) 
and model A 
predictions for the G-dwarf metallicity distribution 
and the [O/Fe] versus [Fe/H] relation. It can be seen that model B
gives essentially the same results as model A.
As discussed in CMG97, the G-dwarf metallicity 
distribution constitutes the most tight constraint for chemical evolution
models available at present. 
The G-dwarf metallicity distribution is representative of the 
history of the chemical
enrichment of the Galaxy, since these stars have lifetimes larger
than or equal to the age of the Galaxy.

Model B is also in agreement with the other available constraints
in the solar vicinity. In Table 2 we present the results of model B
compared with the observational constraints and with the model A
predictions. 
It can be seen that model B results do not differ much from the ones
of model A concerning the observed quantities reported in Table 2.
There are some differences in the predicted ratio of metal-poor and
metal-rich stars and in the $\Delta Y/\Delta Z$ ratio. Model B predictions
are in better agreement for both, although the $\Delta Y/ \Delta Z$ is still
lower than the observed value (Pagel et al. 1992, Chiappini \& Maciel 1994). 
Model B also predicts a value for the relative number of disk and halo stars
which is in better agreement with the observed 
value suggested by Pagel \& Patchett (1975) of 3\%. A higher value
($\simeq 10\%$)
was suggested by Matteucci et al. (1990) who estimated that the number of 
stars contained in a cylinder around the solar neighbourhood with practically
an infinite height above the Galactic plane, which is the actual quantity
predicted by the chemical evolution models, could be a factor 3-4 larger
than the one estimated by Pagel \& Patchett (1975). The range in model predictions
for this constraint is due to different assumptions for the halo phase duration
(1 or 0.8 Gyr).
Model B, however, predicts a ratio between the type II and I supernova rates
which is lower than the observed one. One of the consequences of this can
be seen in Table 3. This Table shows
the solar abundances (by mass) predicted by both models at the time of 
formation of the sun, e.g. 4.5 Gyrs ago,
and the observed solar ones.   
It can be noted that Model B predicts systematically lower values for the 
products of massive star nucleosynthesis (O, Ne, Si, Mg)
and higher ones for the elements
which are mainly produced by low and intermediate mass stars (Fe, N, C)
when comparing with model A results. 
This is due to the fact that the IMF adopted here slightly underestimates 
the number of massive stars. 

As it is well known, the oxygen abundance in the interstellar medium and young stars
in the Orion nebula seems to be smaller by a factor of 2 than the solar one.
Thus it is not clear whether the solar composition should be considered
as representative of the local ISM 4.5 Gyrs ago. In fact, as suggested
by Wielen \& Wilson (1997), the sun could have been born 
at a galactocentric distance
which is roughly 2 kpc smaller than its present one.
Given the uncertainties involved, we can consider that observations
and model predictions are in agreement inside a factor 2 of difference.
This is the case for almost all the elements in Table 3, for models A and B.
However, both models do not predict very well the abundances of 
$^{3}$He and Mg.
As discussed in CMG97 the $^{3}$He is a problem
for almost all the chemical evolution models (see Tosi 1996). For the 
Mg this can be attributed to the lower yield predicted for this element
by Woosley \& Weaver (1995) (see Chiappini et al. 1999).

The good agreement between model B and the observed properties of the Milky Way
at the solar vicinity is due to
the fact that the PNJ IMF, when applied to our
Galaxy, does not vary much over most of the galactic lifetime which is a consequence
of our assumptions of constant molecular cloud temperature during the galactic evolution. In fact, if the cloud temperature and density have varied
strongly over the history of the Galaxy, the predicted IMF will also vary
more and this would certainly worst the agreement with the observational
constraints considered here.

On the other hand, the other two IMFs (models D and E) vary substantially over the entire Galactic 
evolution. As a consequence of that, these IMFs do not give a good agreement with the solar vicinity observational constraints. Figure~\ref{fig5}
 shows the predicted G-dwarf metallicity distribution for models D and E compared with models A and B. Those models
predict too few metal-poor stars and a higher solar metallicity peak than the observed one.
The predicted solar abundances are also not in agreement with the observed ones (see Table 3).
It is worth noting that here we adopted a long timescale for the formation of the solar vicinity
(8 Gyrs) and that a better agreement with the G-dwarf observed metallicity distribution could in principle
be achieved by adopting these variable IMFs in a closed
box model scenario. 
However, as already shown by Scully et al. (1986) and Matteucci \& Tornamb\'e (1985), models
that adopt a shorter infall timescale for the solar vicinity formation do not give good
agreement with all the other observed properties. Moreover, in a close box model the 
predicted gradients along the disk would be essentially flat.

From the comparison with the solar vicinity observational constraints we can
confirm the result by Matteucci \& Tornamb\'e (1985) that only a constant IMF or an IMF 
that varied only at early times can be in agreement with the solar vicinity properties.
 
\subsection{The Galactic Disk}

As stressed by Tosi (1996) it is of fundamental importance to apply the chemical
evolution models not only to the solar vicinity but also to the whole Galaxy. She showed
that it is possible to find models in good agreement with the solar vicinity constraints
that do not fit the radial profiles observed in the galactic disk, namely, the abundance
gradients and the radial gas distribution and that the main differences between the 
various models in the 
literature concern the predictions for the abundance gradient evolution. 
They can be seen as the most
promising constraints to galactic chemical evolution models.

Figure~\ref{fig6} shows the behavior of model B IMF at different galactocentric
distances.
This figure shows that the radial variation of the adopted IMF in model B is
important only for low mass stars ($M \leq 0.5 M_{\odot}$), and predicts
a higher number of them towards the galactic center, where the metallicity
is higher. A direct consequence of this fact is the predicted flatter
gradient with respect to the ones presented by CMG97 (Model
A). Figure~\ref{fig7} shows the oxygen abundance gradient for 3 different models: model
A (CMG97), model B (adopting the PNJ IMF) and
model C (same as B but with star formation efficiency increasing with
decreasing galactocentric distance). Model B predicts a flat gradient between 6 and 10 kpc,
and a small negative one in the inner part of the Galaxy ($ R < 6 kpc $). 
Model C, which adopts an
increasing star formation efficiency $\nu$ towards the galactic center, predicts
a steeper gradient inside the solar circle. In this
case a the bimodal abundance gradient, steeper in the inner part and
flatter outside, similar to the one found by CMG97,
is recovered. 
However, as can be seen in Figure~\ref{fig8}, a model with a higher star formation
efficiency 
in the central parts (Model C) consumes the gas very fast thus
reaching the threshold density value very soon. As a consequence, model C 
predicts a flat gas density radial profile
at variance with observations, whereas Model B predicts a gas density distribution
very similar to the one predicted by model A and in better agreement with observations. 

Figure~\ref{fig8} also shows models D and E predictions for the radial gas profile.
From this figure it can be seen that 
the only two models in good agreement
with the observed radial profile are models A and B. However, as shown before (see Figure~\ref{fig7}),
model B predicts even flatter abundance gradients than model A. 
Models D and E
predict a flatter radial gas profile than the observed one but in better agreement with the observations
than the one predicted by model C. Moreover, the oxygen abundance gradient predicted by models
D and E (see Figure~\ref{fig9}) has a slope which is similar to the one obtained with model A.
In any case, those two models do not give predictions in agreeement with the solar vicinity
constraint as already shown in the previous section.

This is a very important result showing that only model A, with a constant IMF, is in agreement
simultaneously with the solar vicinity and the disk observational constraints. The 
abundance gradients predicted by model A, although slightly flatter than the observed, are steeper 
inside than in the outer parts of the 
galactic disk and also steepen with time, in agreement with the recent results by Maciel
\& Quireza (1999). 

Figure~\ref{fig10} shows the oxygen gradient evolution in time for model B.
The present model (B), with a variable IMF, predicts a more complex
evolution for the radial abundance gradients than model A (see Figure~\ref{fig11}).
In this case, although the general behavior is the same as in A (a gradient
which becomes more negative in the inner parts and is roughly constant
in the outer ones), there is an intermediate region where the gradient
is zero today and was positive at early epochs. 

Another approach was recently proposed by Carigi (1996).
She adopted a simple parameterized IMF which depends on the metallicity,  
taking into account the suggestion by Silk (1995) that at low-metallicities
the IMF should be more biased towards the formation of low-mass stars than
the Scalo one. This is exactly the opposite situation with respect to the IMFs
presented here. She took the IMF from Kroupa et al. (1993) and adopted a 
metallicity dependence on the slope of the their IMF for low-mass stars ($M \leq 0.5
 M_{\odot} $). Her IMF contains an explicit dependence on the oxygen abundance
and predicts that the number of very low mass stars decreases as oxygen
increases. She applies this IMF to her chemical evolution model (one
infall episode) and shows that such a model produces radial abundance
gradients which steepen with time (at variance with her previous models
but in agreement with the CMG97 model) but fails, as
expected, in 
reproducing the G-dwarf metallicity distribution, predicting two many low
metallicity stars (the G-dwarf problem).

\section{Discussion and Conclusions}

In this work we adopted a new IMF proposed by PNJ  and
applied it to the {\it Two Infall Chemical Evolution Model} 
(CMG97) and other two strongly variable IMFs already suggested in the 
literature (Matteucci \& Tornamb\'e 1985 and Scully et al. 1996).
Those IMFs all predict a higher number of massive stars at the beginning 
of the Galaxy evolution and in the outermost regions of the Galactic disk.

The two-infall model coupled with the PNJ IMF
reproduces most observational constraints
for the solar vicinity, namely, the G-dwarf metallicity distribution,
the [$\alpha$/Fe] versus [Fe/H] behavior, the solar abundances, 
the gas density, etc. and this is a consequence of the mild variation of such
IMF when applied to our chemical evolution model under the hypothesis of constant
molecular cloud temperature and density, allowing only for a 
variation in the velocity of dispersion throughout the galaxy evolution.

However, the predicted gradient when adopting this IMF (model B) is flatter than the one obtained 
with a constant IMF at variance with observations.
This is due to the fact that the peak mass of the PNJ IMF is shifted towards higher mass values during the 
early phases of the Galaxy evolution. This, combined with an inside-out
picture for the thin disk formation, leads to a flat abundance gradient.
In this scenario the outer parts of the thin disk
are less evolved than the central ones
and hence the IMF there is biased towards higher mass stars.
In order to obtain a steeper gradient, a stronger dependence
of the star formation rate on the galactocentric distance should be invoked. 
This, however, leads to a flat gas density profile at variance with observations. 

This fact suggests that models assuming a star formation
rate which is a strong function of the radial galactocentric distance
(e.g. Prantzos \& Aubert 1995) must be also
confronted with the observed radial density profile and the abundance
gradients. A strong radial variation for the star formation rate was
adopted by Prantzos \& Aubert (1995) but they 
also adopted a mild dependence of the SFR 
on the gas density namely, a smaller $k$ value than the one adopted here. 
In this case, even
with a SFR which is a strong function of the galactocentric
distance they could obtain a gas density profile
which strongly decreases outside the solar circle. 
In fact, as discussed by Matteucci \& Fran\c cois (1989), the total surface
gas density along the galactic disk crucially depends on the exponent
of the star formation law, $k$, and on the parameter $\tau_D(r)$.
The best value for $k$, as showed
by CMG97, who obtained a good fit of the recent
observed G-dwarf metallicity distribution (Rocha-Pinto \& Maciel 1996)
is larger than
that used by Prantzos \& Aubert (see Matteucci 1996 for a discussion)
and is coupled with a timescale for the thin disk formation, at solar vicinity,
of 8 Gyr. These are the two most important parameters of the models and
are constrained into a very narrow range of values (see CMG97).

When adopting strongly variable IMFs along with the two-infall model we can obtain
better gradients but a worst agreement with other observational constraints in the 
solar vicinity and the radial gas profile.
Recently, Martinelli \& Matteucci (1999) proposed a method to solve the G-dwarf problem 
in a closed box model with
a time-dependent IMF. The IMFs they proposed were tested by using the two-infall model of CMG97 and 
the main conclusion reached was that 
an IMF similar to the one proposed by Larson (1998) fails in reproducing
the [O/Fe] versus [Fe/H] relation observed in the solar vicinity, even if it gives
a good fit of the G-dwarf metallicity distribution, because of the strong time dependence
of the number of type II SNe at early times.
Their study also showed that a better fit to the abundance ratios is obtained when a constant
IMF is adopted.

Contrary to the suggestion by Larson (1998) that it now appear that most of the known
intergalactic gas has been heated and enriched in heavy elements by the effects of early
star formation in galaxies, we do not think this was the case at least for galaxies like the Milky Way.
For the Milky Way, in fact,
a model with a constant IMF and a long timescale
for the formation of the solar vicinity is still the best way to solve the G-dwarf problem.

It is interesting to note that
not only from the theoretical side we can conclude
that models with a longer timescale for the galactic disk formation are better in reproducing
the observational constraints, but also from the observational side there is now evidence
for infall. 
In a recent paper Blitz et al. (1999) suggest that the
high velocity clouds represent infall of intergalactic medium onto the Local Group and those clouds 
could also represent the building blocks from which the Local Group was assembled and they continued
to fuel the star formation on the disk of the Milky Way.
From a simulation of the dynamics of the HVCs the authors estimate that the accretion 
of HVCs by the Milky Way was rapid early on, reaching a peak of about 30 times the current
rate (estimated as being of the order of 7.5 M$_{\odot}$/yr$^{-1}$) withing the first 1 Gyr
after the beginning of the simulation and that half of the mass accreted by the MW falls onto it 
during the first 2 Gyrs. The authors also show that the few available measurements from absorption
lines give metallicities much lower than solar in agreement with the idea that those clouds
do not represent gas that left the disk through galactic fountains as has often been suggested in the
literature (Bregman 1980). Those simulations are still very uncertain mainly because of the uncertainty 
on the distance to HVCs. Better data on HVCs are fundamental to confirm or discard the existence of 
infall onto the Milky Way.

Another argument in favor of models that adopt infall for the formation of the disk of the Milky Way
is also shown by Roche et al. (1998). Those authors estimate that galaxies at 2$<$ redshift $<$ 3.5
show 2.79 $\pm$ 0.31 mag of surface brightness evolution relative to those at redshift $<$ 0.35, which
is significantly larger than the luminosity evolution over this redshift range. They suggest that this
can be explained by a size and luminosity evolution model, in which the outer regions of spiral galaxies
form later and with a longer time-scale than the inner regions, causing the half-light radius to increase 
with time. To do this kind of models those authors adopted the $\tau_D(r)$ proposed by CMG97 as 
corresponding to the SFR timescales at different radius of a spiral galaxy.
 
To summarize, 
the main results of the present paper are:
\par\noindent
a) We tested the IMF proposed by PNJ in a model for the chemical 
evolution of the Milky Way (CMG97) and we showed that this IMF
gives good agreement with the observed properties of the solar vicinity.
\par\noindent
b) We concluded that such an agreement is due to the fact that this IMF when applied
to the two-infall model 
shows a time variation that is important only 
in the early phases of Galactic evolution. This in turn is due to the 
simplifying assumptions adopted here, like neglecting the dependence of the IMF on the gas temperature which would produce more sharply varying IMF. In these early phases the IMF is biased towards massive stars.
\par\noindent
c) the PNJ IMF combined with the inside-out picture for the thin disk
formation predicts a gradient flatter than the one predicted by a model which adopts a constant
IMF. This situation cannot be reversed by changing the SFR because in this case
the abundance gradient is recovered but the gas density profile is destroyed.
To better constrain the radial dependence of some of the adopted
parameters, like $\tau(r)$ or $\nu_D(r)$, more reliable observations are needed
especially for the gas, star formation and abundances radial profile.
\par\noindent
d) Models which adopt IMFs strongly dependent on metallicity, thus simulating a dependence
also on the gas temperature, are not
in agreement with the most important observational constraints of the solar vicinity
and predict radial gas profiles at variance with observations, therefore they should
be rejected.
\par\noindent
e) We conclude that a constant IMF and the assumption of a continuous infall onto the
galactic disk is still the best way to explain the observational constraints in the Milky Way
including the G-dwarf metallicity distribution.

\acknowledgements
\noindent{\bf acknowledgments} 
The authors want to thank the SISSA institute and the Department
of Astronomy of the University of Trieste for their kind hospitality.
This work was partially supported by CNPq and FAPESP, Brazil. C. C. and F. M. express their
gratitute to the STScI for hospitality during their visit, when the final version of this paper
was completed and thank Dr. C. Leitherer and Dr. N. Panagia for the invitation. 
We also thank the referee, J. Scalo, for very usefull comments and suggestions.

\newpage

\figcaption[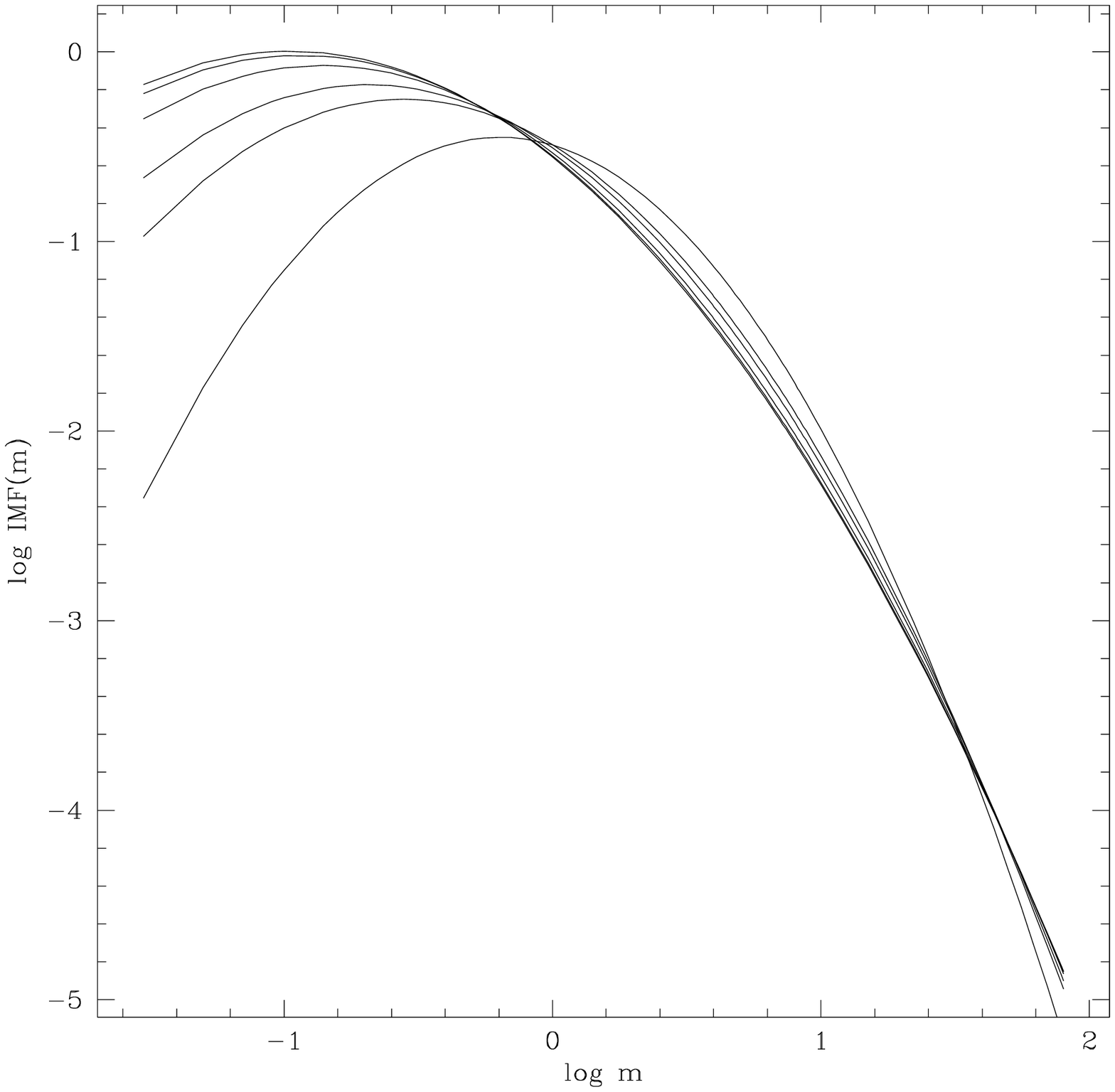]{The PNJ IMF, as function of time, for the solar
vicinity for t = 15, 10, 5, 2, 0.6 and 0.02 Gyrs (from top to bottom).
\label{fig1a}}

\figcaption[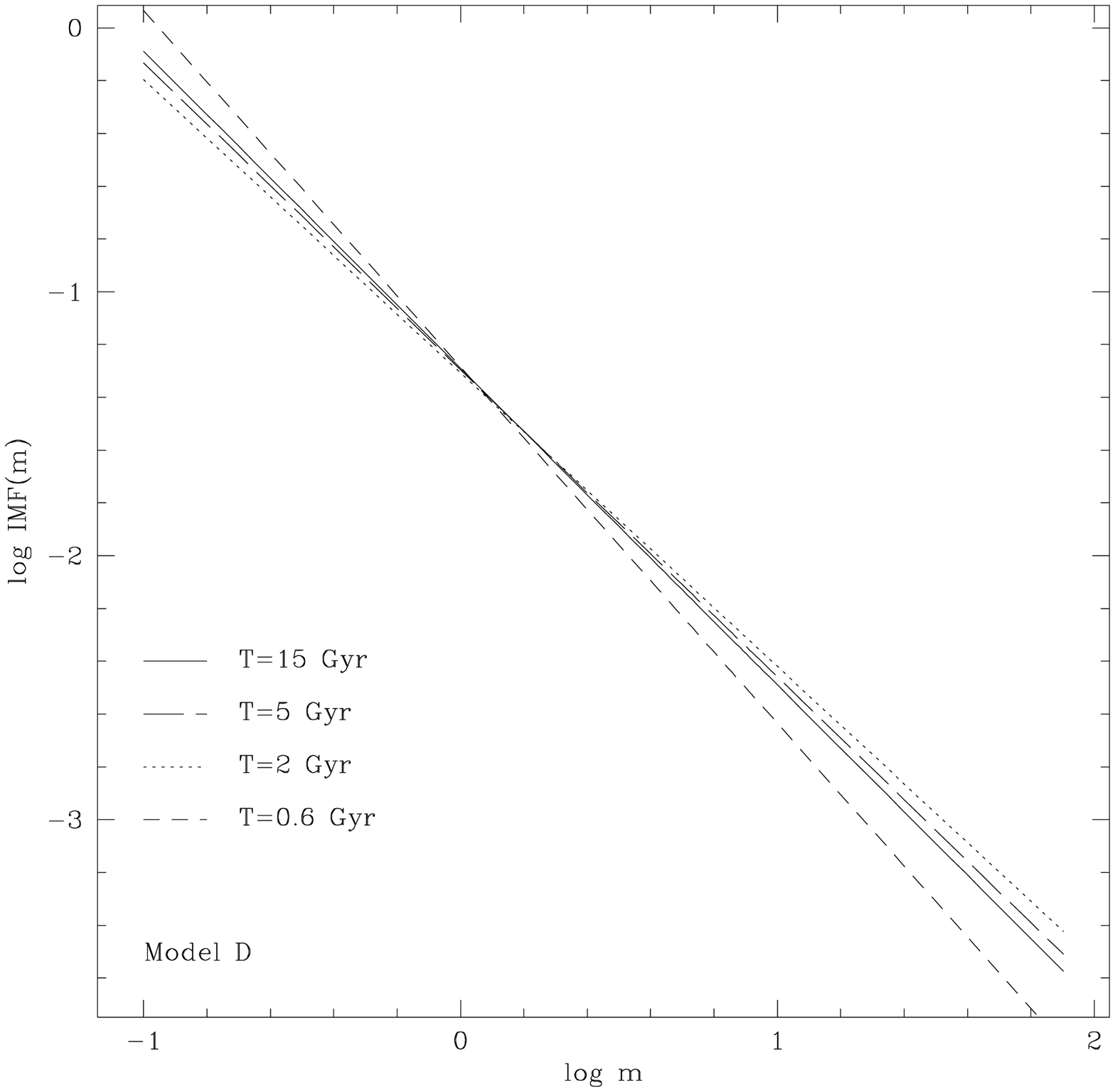]{The model D IMF as a function of time for the solar vicinity
\label{fig1b}}

\figcaption[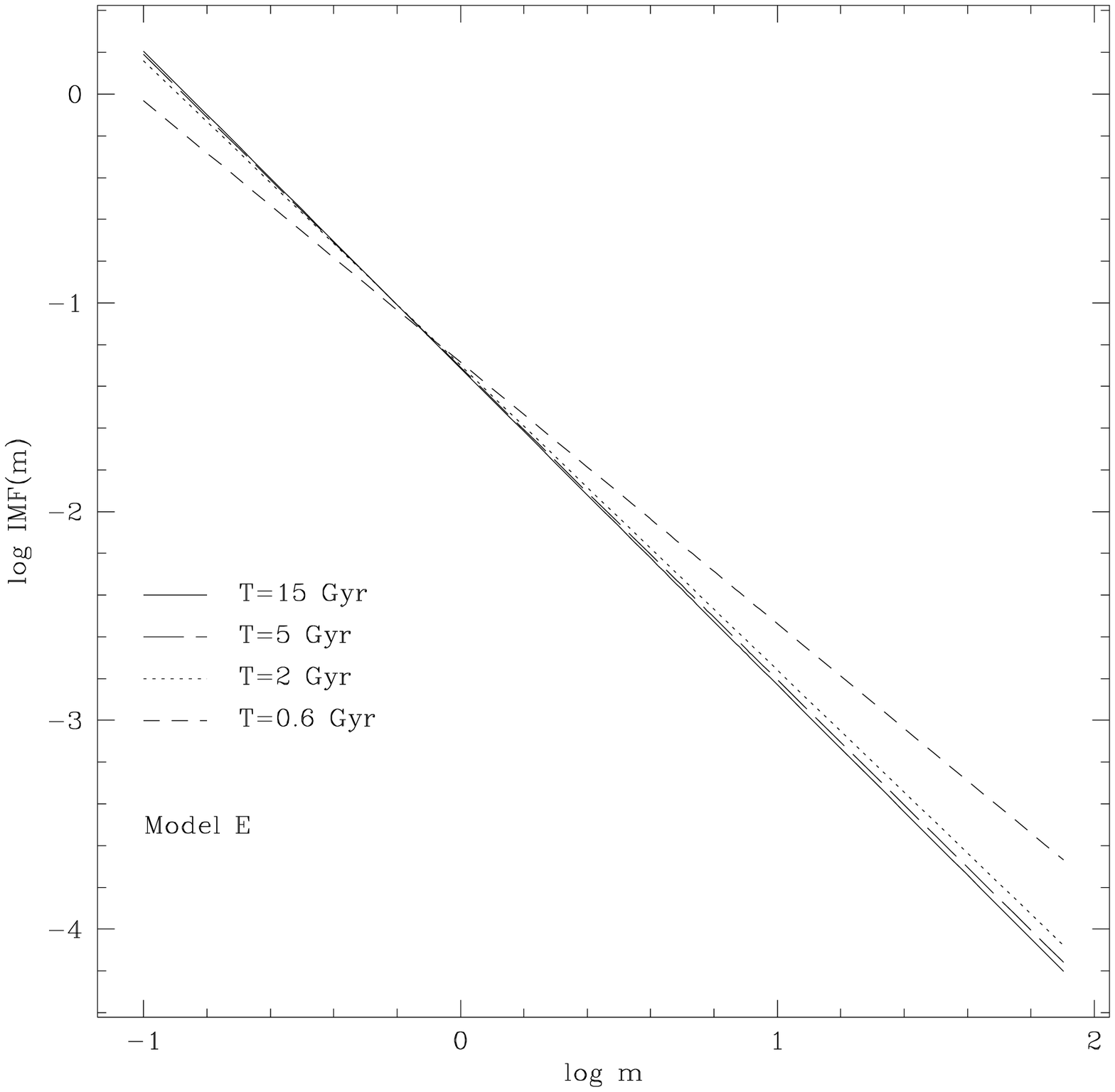]{The same as Figure~\ref{fig1b} for model E
\label{fig1c}}

\figcaption[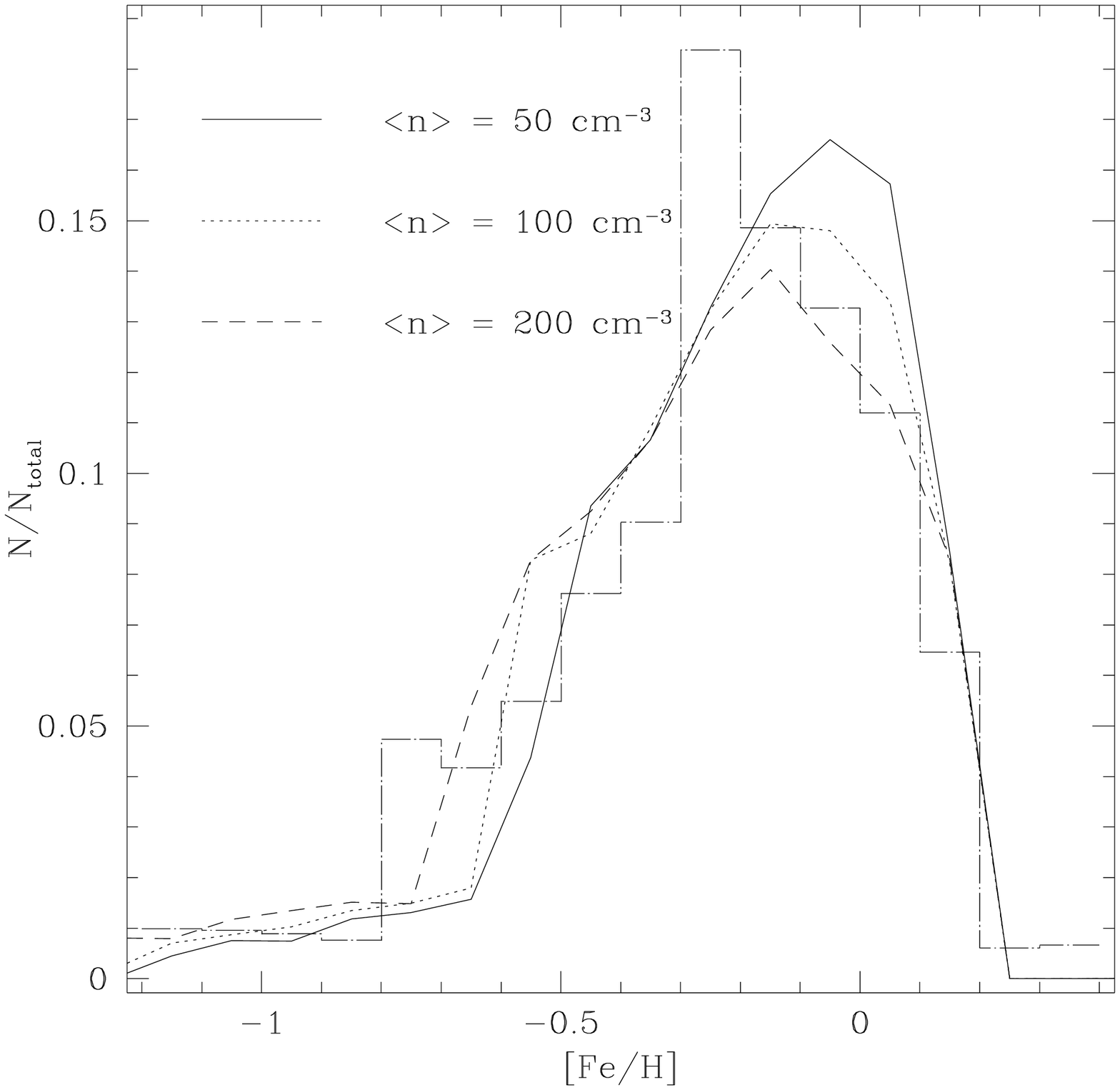]{The effect of the density parameter $<n>$ on the
model B predictions for the G-dwarf metallicity distributions for models
which adopt the variable IMF.
\label{fig2}}

\figcaption[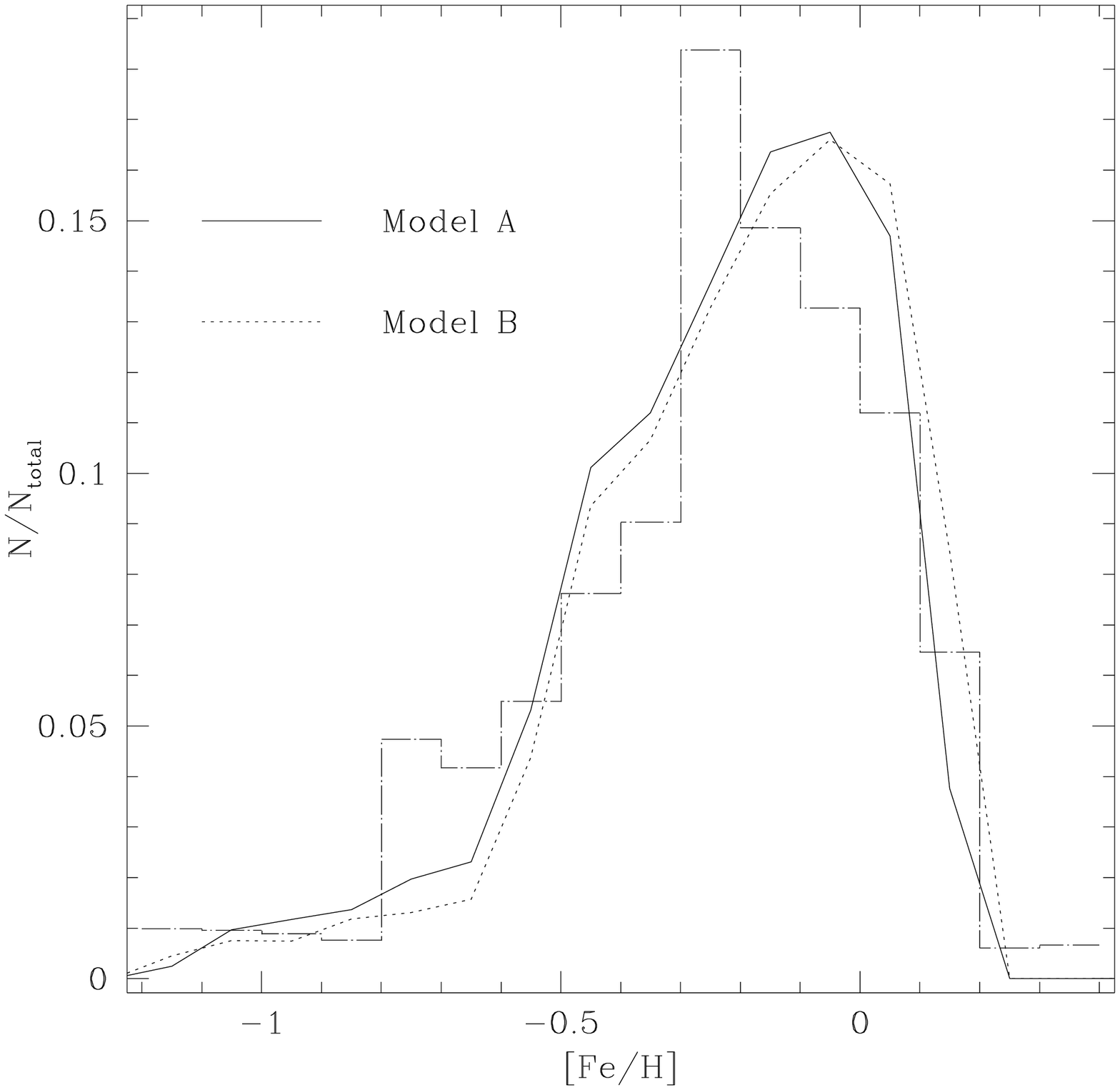]{The G-dwarf metallicity distribution. The data are
from Rocha-Pinto \& Maciel (1996). The lines refer to model predictions
for model A (solid line) and model B (dotted line).
\label{fig3}}

\figcaption[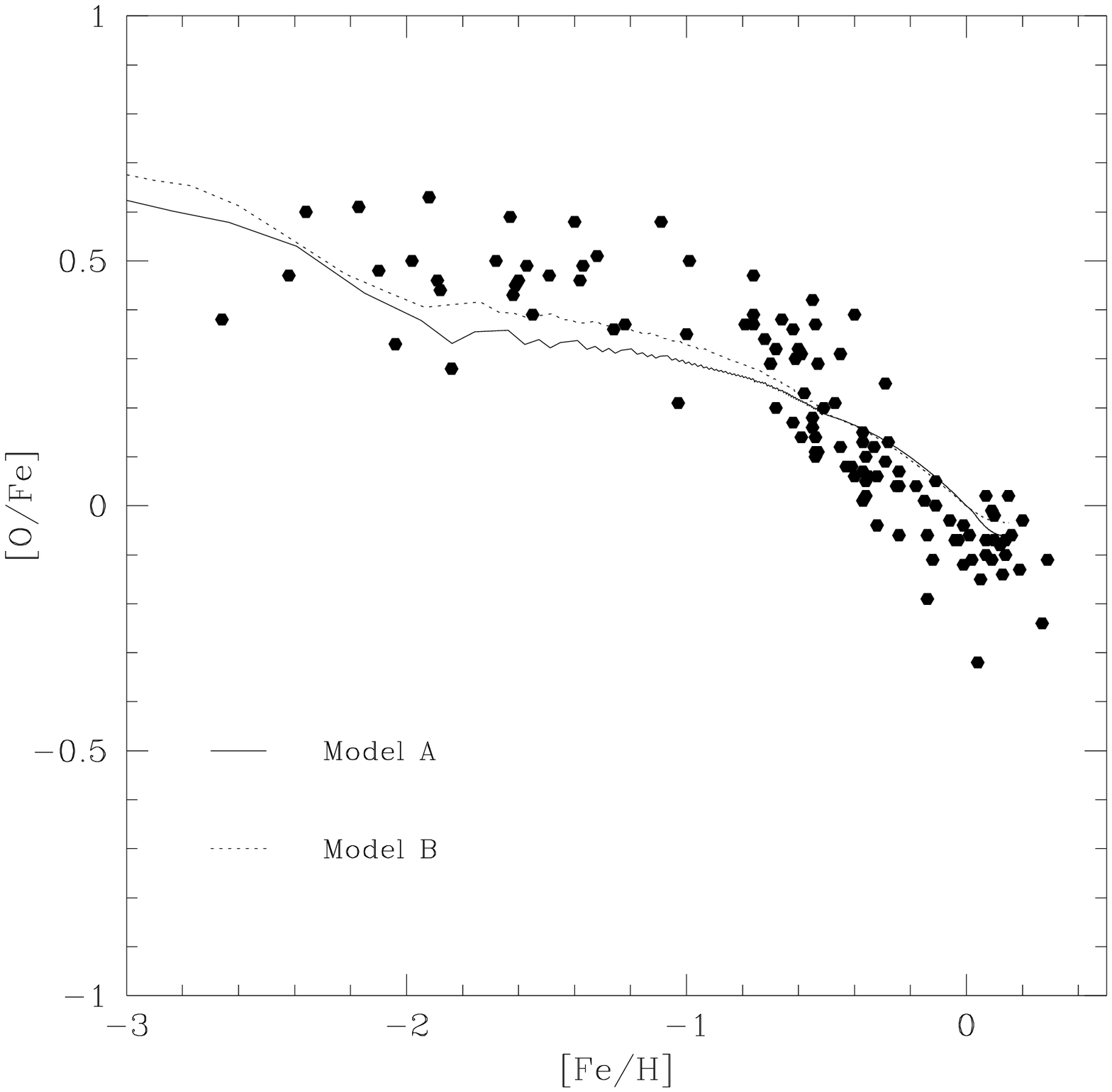]{[O/Fe] versus [Fe/H] behavior. The data were taken from
Gratton et al. (1997). The curves are labeled as in Figure~\ref{fig3}.
\label{fig4}}

\figcaption[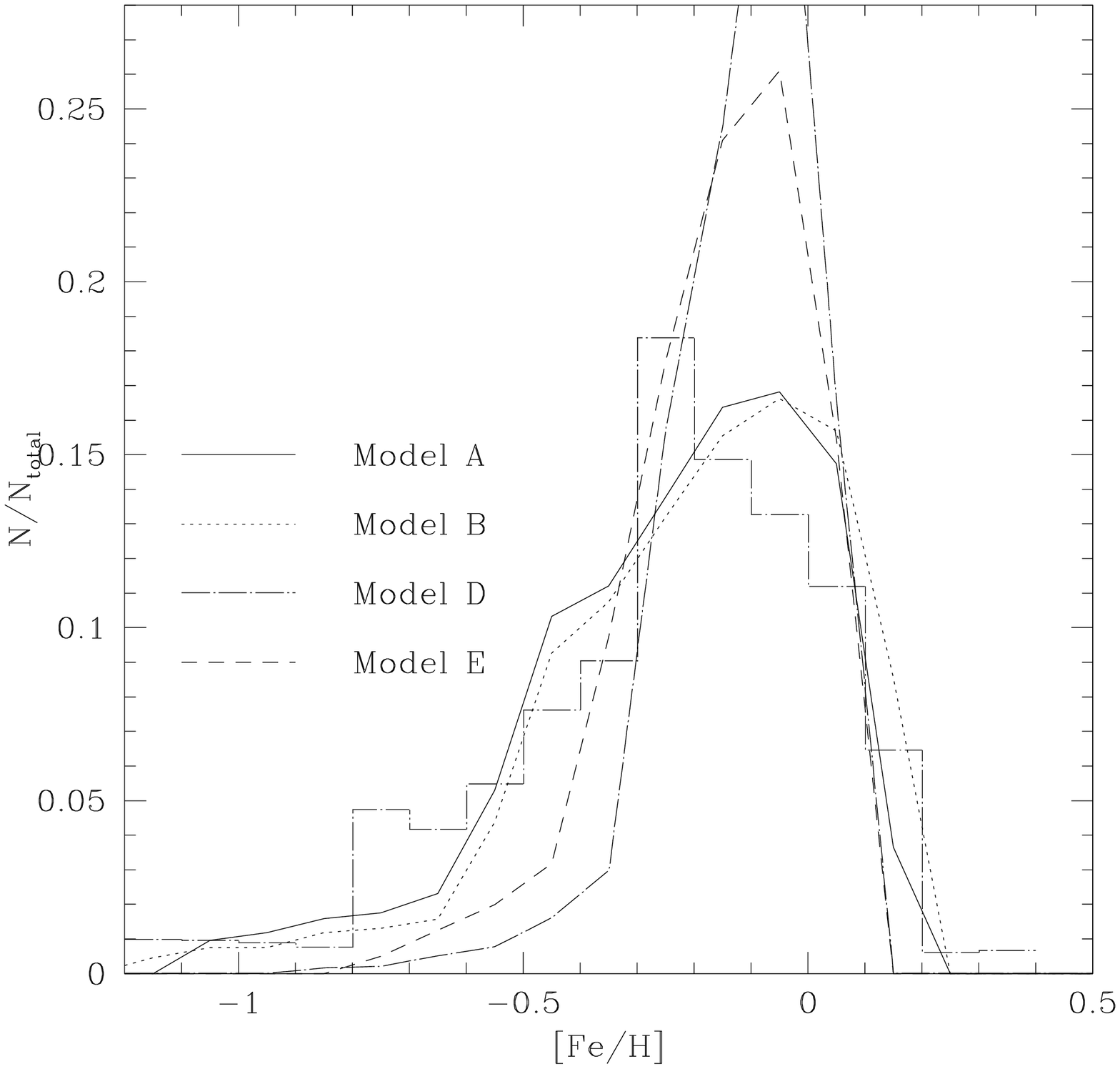]{The G-dwarf metallicity distribution predicted
by models A, B, D and E. The data are
from Rocha-Pinto \& Maciel (1996).
\label{fig5}}

\figcaption[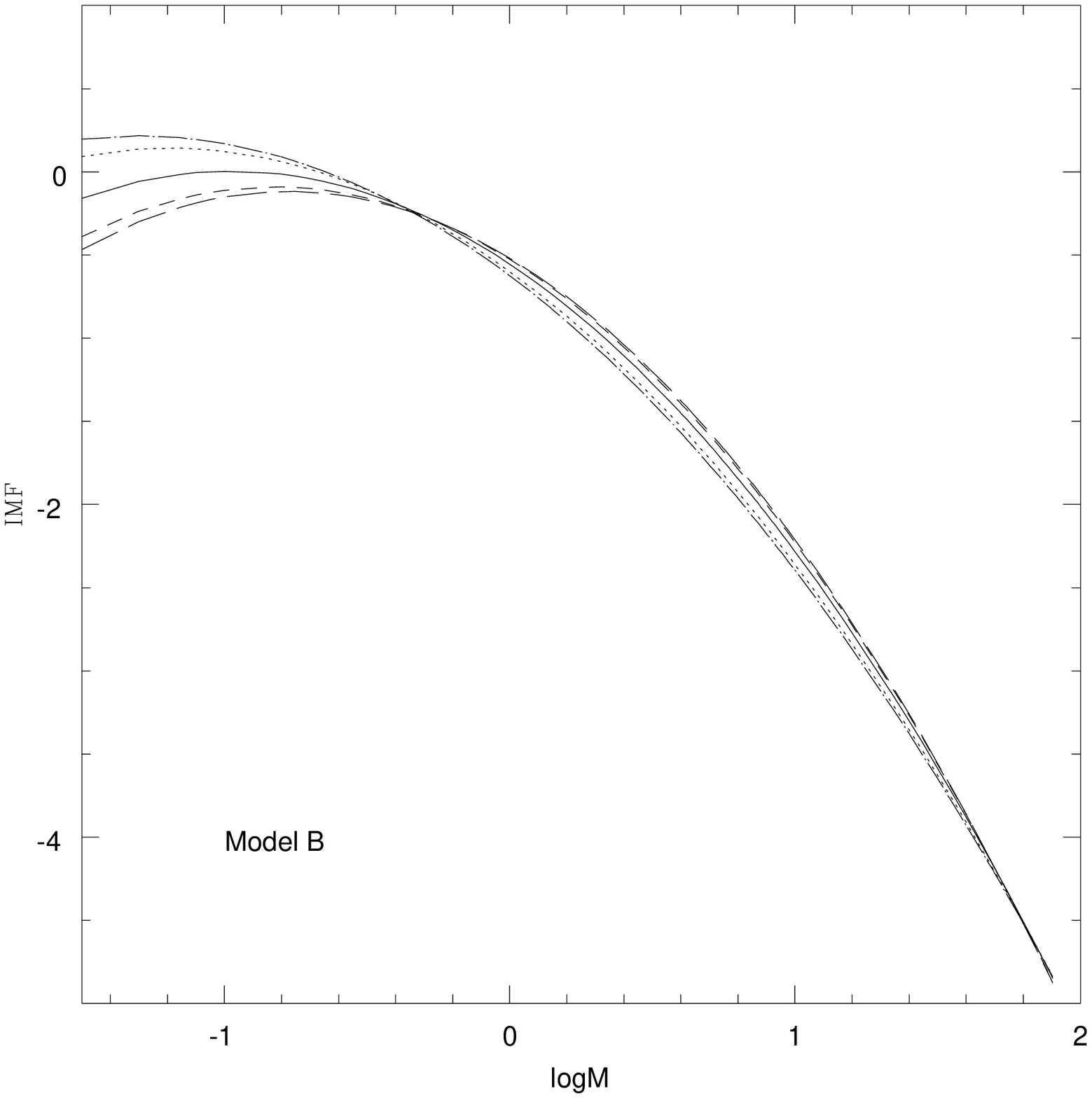]{The radial dependence of the PNJ 
when applied to our Galaxy, at present. 
In our model the solar vicinity corresponds to a galactocentric
distance of 10 kpc. The five curves refer to different distances from the 
galactic center (dot-dashed line: 4 kpc; dotted line: 6 kpc ; 
solid line: solar vicinity ; short-dashed line: 14 kpc and long-dashed
line: 16 kpc).
\label{fig6}}

\figcaption[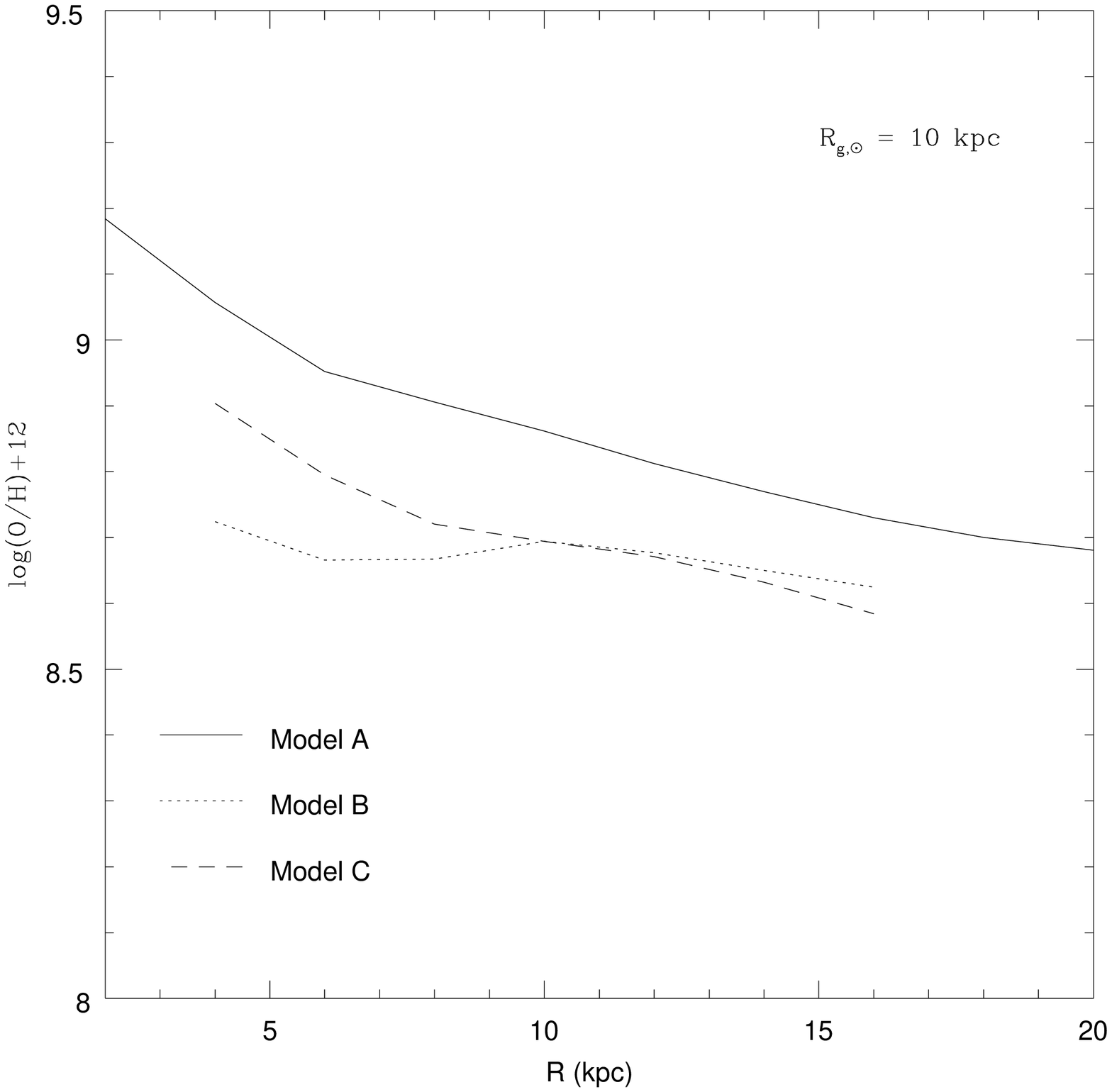]{Oxygen abundance gradient for models A, B and C.
\label{fig7}}

\figcaption[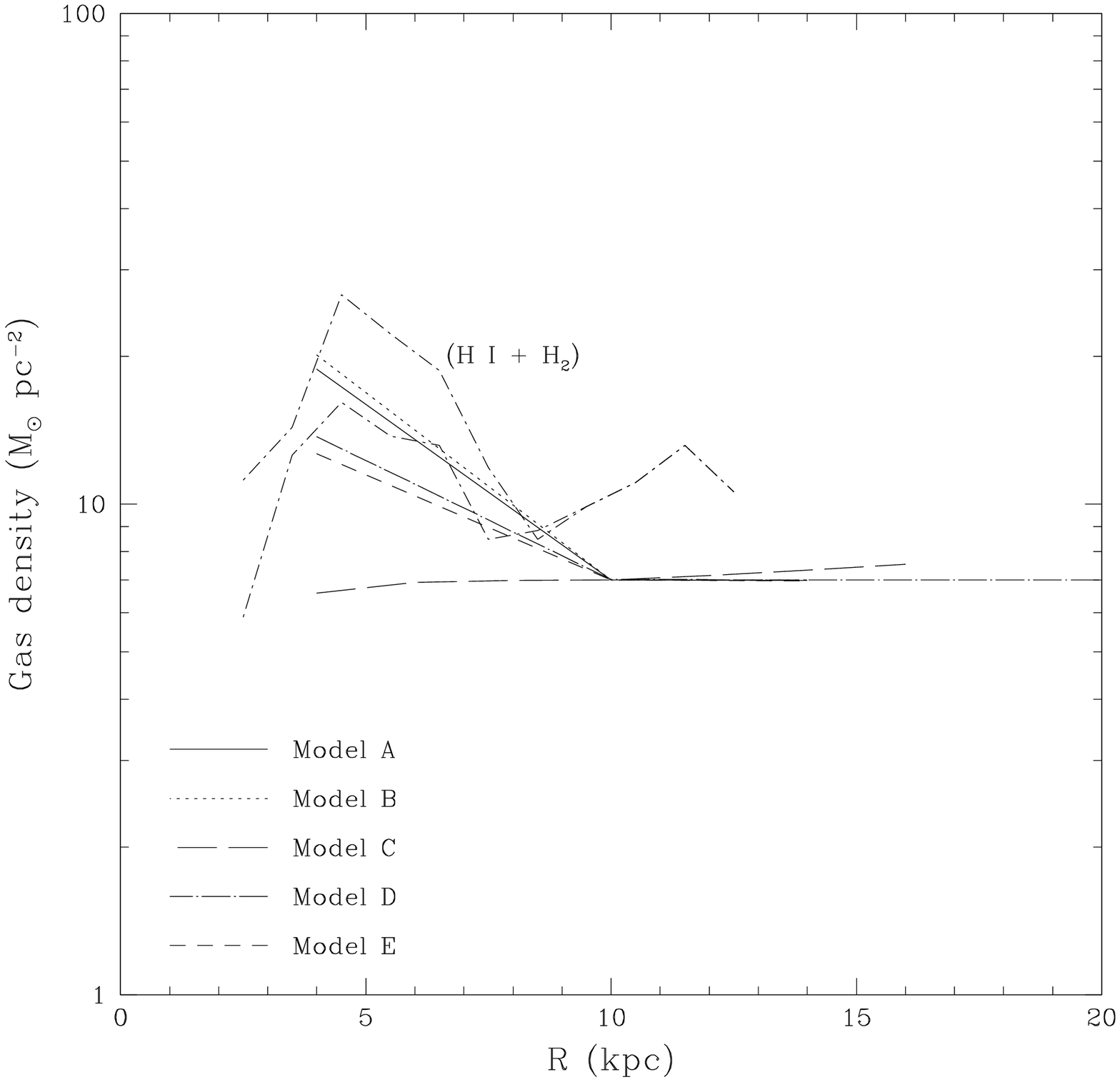]{Total surface gas density distribution given by Rana (1991) (short dashed-long dashed line)
The predictions from all models of Table 1 are also shown.
\label{fig8}}

\figcaption[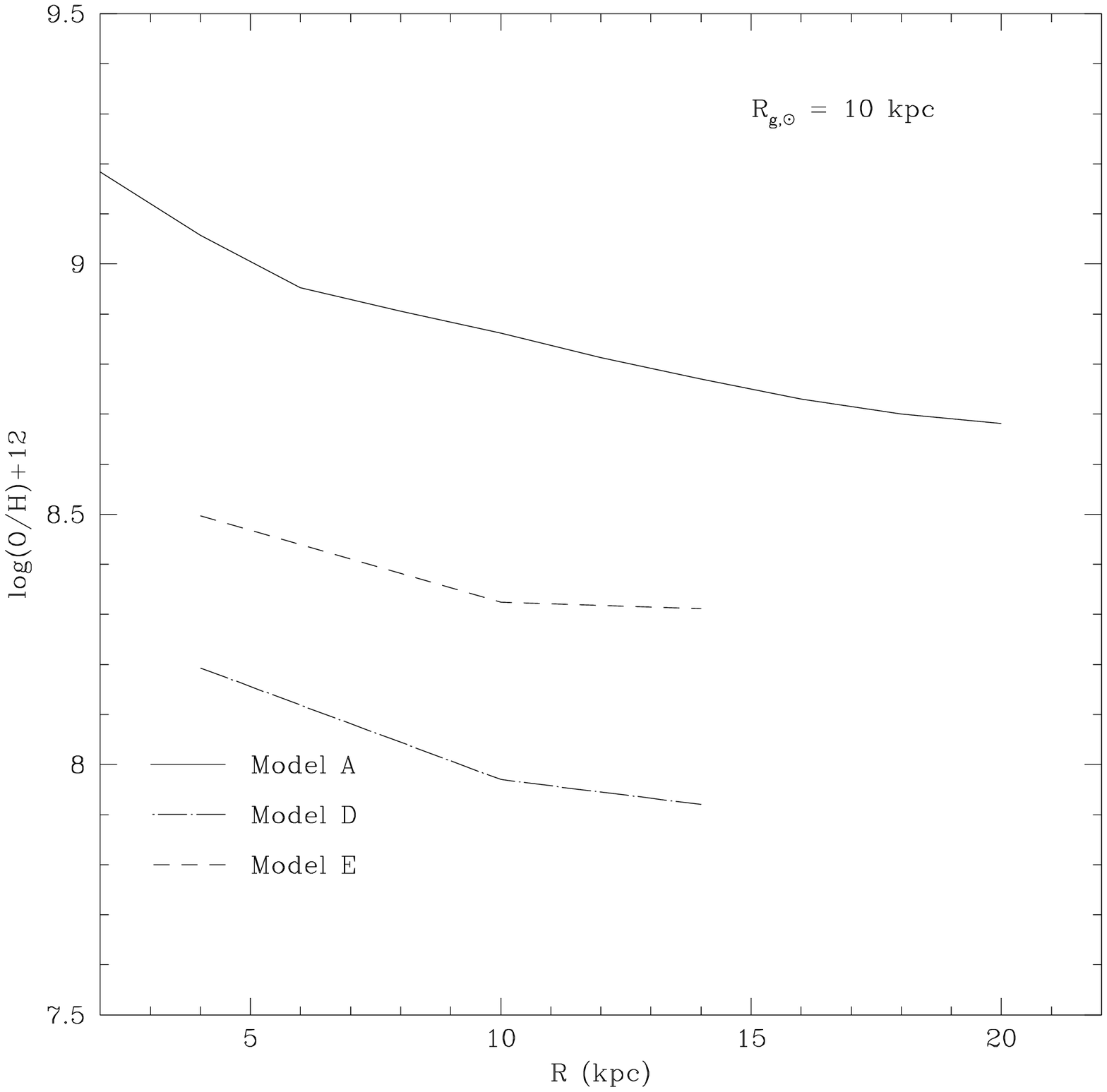]{Oxygen abundance gradient predicted by models 
A, D and E
\label{fig9}}

\figcaption[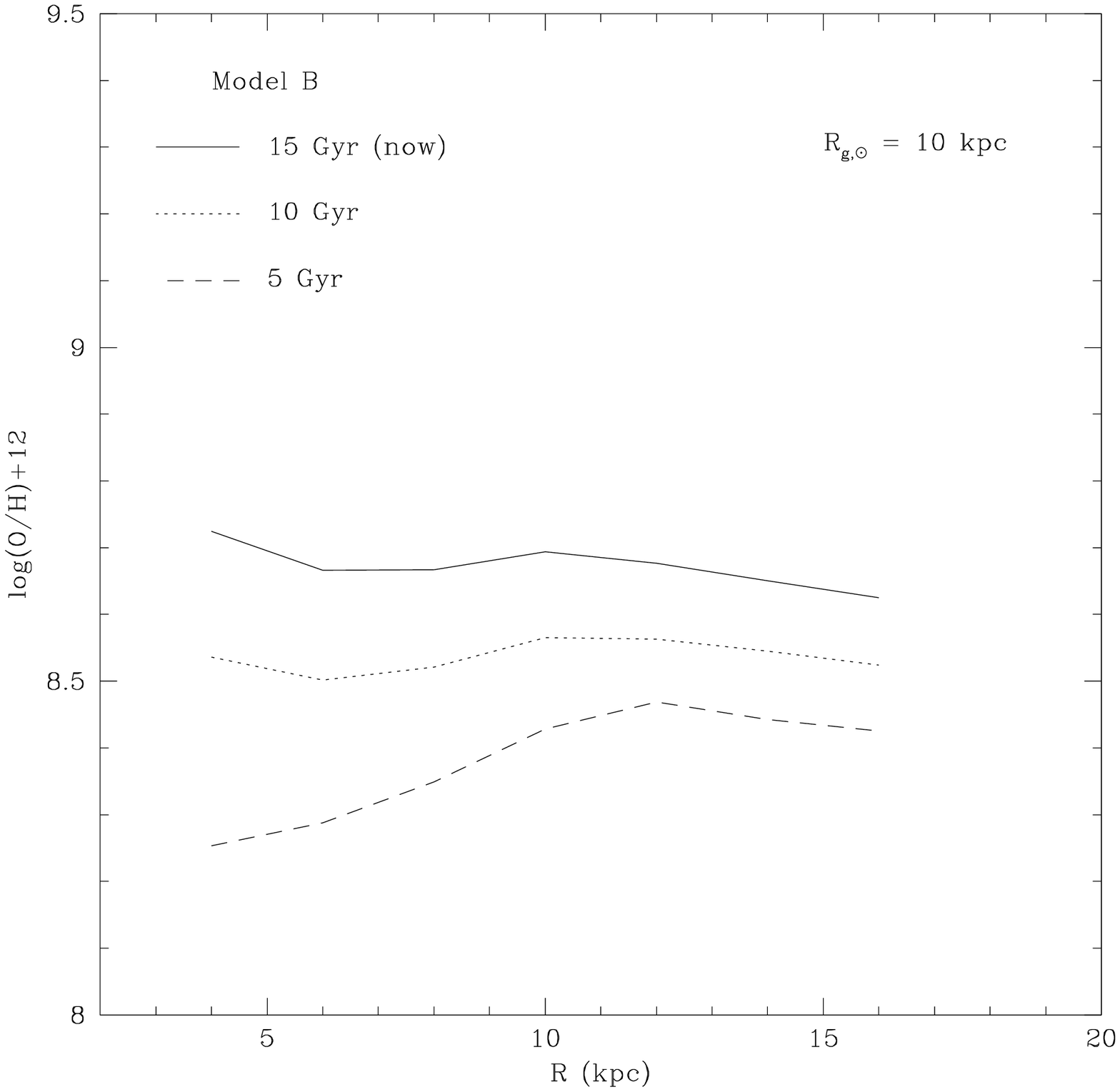]{Evolution of the oxygen abundance gradient for model B.
\label{fig10}}

\tiny
\figcaption[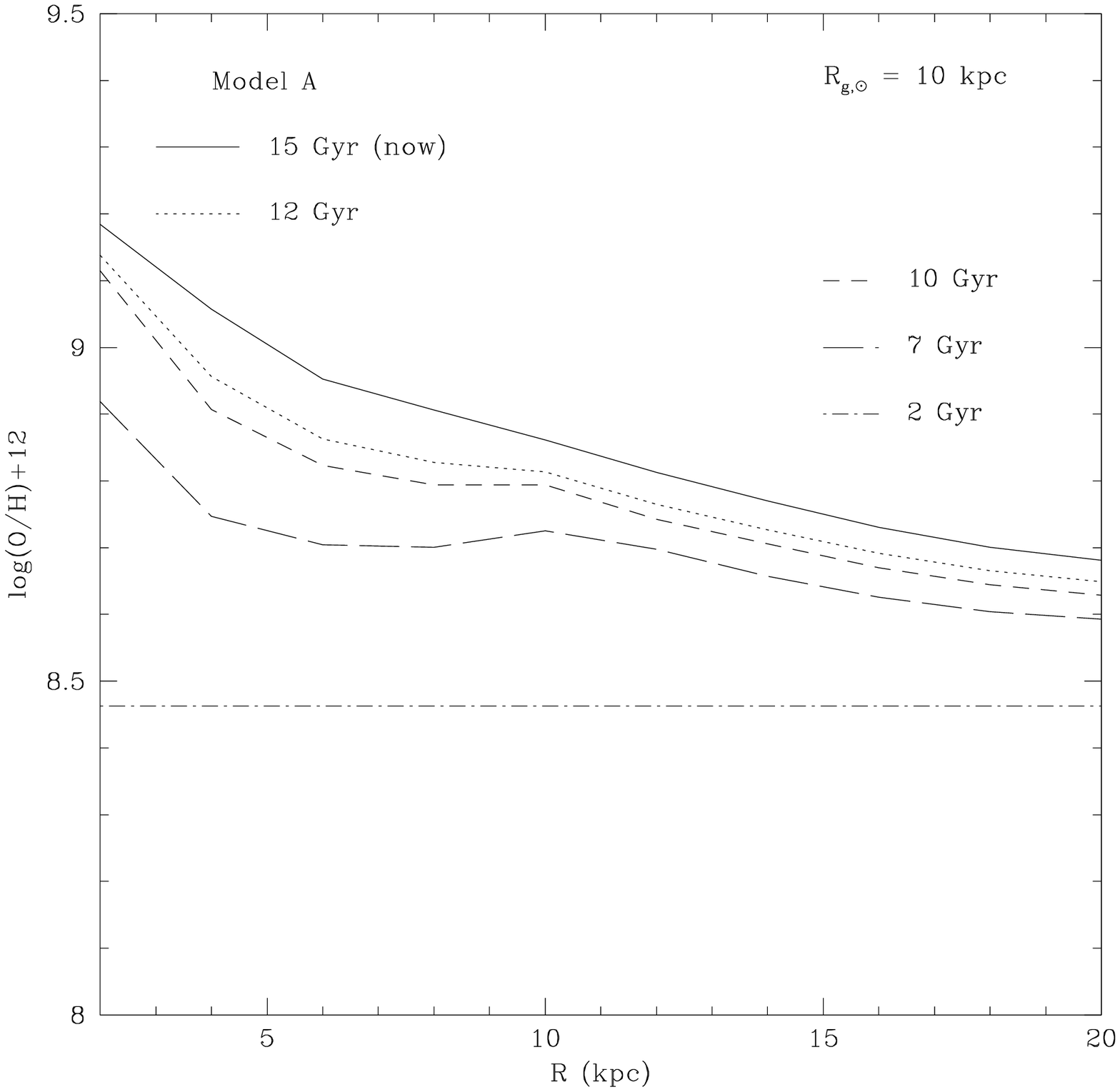]{The same as Figure~\ref{fig10} for model A.
\label{fig11}}
\normalsize
\newpage

\tiny
\begin{table*}
\caption{Input Model Parameters}
\smallskip
\label{}
\begin{tabular}{cccccccr}
\hline
 & & & & & & &\\
Model & ${\tilde\nu}_H$ & $k_H$ & ${\tilde\nu}_D$ 
& $k_D$ & $\tau_H$ & $\tau_D$ & IMF \\
 & ($Gyr^{-1}$) & & ($Gyr^{-1}$) & & ($Gyr$) & ($Gyr$) & \\
\smallskip
 & & & & & & & \\
\hline
& & & & & & & \\
A & 2.0 & 1.5 & 1.0 & 1.5 & 1.0 & 8.0 & Scalo (1986) \\
& & & & & & & \\
B & 2.0 & 1.5 & 1.0 & 1.5 & 1.0 & 8.0 & PNJ\\
& & & & & & & \\
C & 2.0 & 1.5 & $\nu(r)$ $\propto r^{-4}$ & 1.5 & 1.0 & 8.0 & PNJ\\
  &     &     & and 1.0 for $R_{g\odot}$ & & & & \\
& & & & & & & \\
D & 2.0 & 1.5 & 1.0 & 1.5 & 1.0 & 8.0 & Matteucci and Tornamb\'e (1985)\\
& & & & & & & \\
E & 2.0 & 1.5 & 1.0 & 1.5 & 1.0 & 8.0 & Scully et al. (1996)\\
\hline
\end{tabular}
\end{table*}
\normalsize

\newpage

\begin{table}
\caption{Current predicted and observed quantities for the solar
neighbourhood for models A and B}
\smallskip
\label{}
\begin{tabular}{cccr}
\hline
  & Model A & Model B & Observations \\
\hline
Metal-poor/total stars (\%) 
 & 6-13 \% & 2-4 \% & 2 - 10 \% \\
SNII/SNI & 2.7 & 1.4 & 2.4 \\
$\Psi$(R$_{g,\odot}$,t$_{now}$) (M$_{\odot}$ pc$^{-2}$ Gyr$^{-1}$) & 2.64 & 2.65  & 2-10  \\
$\sigma_g$(R$_{g,\odot}$,t$_{now}$) (M$_{\odot}$ pc$^{-2}$) & 7.0 & 7.0 & 6.6 $\pm$ 2.5 \\
$\sigma_g$ / $\sigma_T$ (R$_{g,\odot}$,t$_{now}$) & 0.14 & 0.14 & 0.05-0.20 \\
$\dot{\sigma_{inf}}$(R$_{g,\odot}$,t$_{now}$)(M$_{\odot}$ pc$^{-2}$ Gyr$^{-1}$)  & 1.05 & 1.05 & 1.0 \\
$\Delta$ Y/$\Delta$ Z & 1.63 & 1.96 & 3.5 $ \pm $ 0.7 \\
$\Psi$(R$_{g,\odot}$,t$_{now}$)/$\langle \Psi \rangle$ & $\sim$ 0.7 & $\sim$ 0.7 & 0.18-3.0 \\
X$_2(P)$/X$_2(now)$ & 1.51 & 1.82 & 1.4-11.9 \\
\hline
\end{tabular}
\end{table}

\newpage

\begin{table*}
\caption{Solar Abundances by Mass ($^*$ at 4.5 Gyrs ago)}
\smallskip
\label{}
\begin{tabular}{cccccr}
\hline
&&&&& \\
Element & $^*$Model A & $^*$Model B & $^*$Model D & $^*$Model E & Observations \\
 &  &  & & & Anders and Grevesse (1989) \\
 & & & & &\\
\hline
 & & & & &\\
H & 0.731 & 0.730 & 0.736 & 0.758 & 0.702  \\
D & 4.63 (-5) & 4.10 (-5) & 5.27 (-5) & 5.72 (-5) & 4.80 (-5)  \\
$^{3}$He & 10.05 (-5) & 14.0 (-5) & 5.48 (-5) & 5.31 (-5) & 2.93 (-5) \\
$^{4}$He & 2.55 (-1) & 2.57 (-1) & 2.50 (-1) & 2.37 (-1) & 2.75 (-1) \\
$^{12}$C & 1.83 (-3) & 2.59 (-3) & 1.17 (-3) & 0.49 (-3) & 3.03 (-3) \\
$^{16}$O & 7.29 (-3) & 4.39 (-3) & 8.26 (-3) & 2.45 (-3) & 9.59 (-3) \\
$^{14}$N & 1.39 (-3) & 1.79 (-3) & 0.96 (-3) & 0.27 (-3) & 1.11 (-3) \\
$^{13}$C & 4.76 (-5) & 7.40 (-5) & 2.51 (-5) & 0.99 (-5) & 3.65 (-5) \\
Ne & 0.94 (-3) & 0.53 (-3) & 1.12 (-3) & 0.32 (-3) & 1.62 (-3) \\
Mg & 2.48 (-4) & 1.55 (-4) & 2.76 (-4) & 0.83 (-4) & 5.15 (-4) \\
Si & 7.04 (-4) & 6.71 (-4) & 6.00 (-4) & 2.15 (-4) & 7.11 (-4) \\
S & 3.07 (-4) & 3.11 (-4) & 2.51 (-4) & 0.93 (-4) & 4.18 (-4) \\
Ca & 3.95 (-5) & 4.16 (-5) & 3.13 (-5) & 1.18 (-5) & 6.20 (-5) \\
Fe & 1.37 (-3) & 1.79 (-3) & 0.92 (-3) & 0.39 (-3) & 1.27 (-3) \\
Cu & 8.20 (-7) & 9.29 (-7) & 5.96 (-7) & 1.58 (-7) & 8.40 (-7) \\
Zn & 2.45 (-6) & 2.93 (-6) & 1.71 (-6) & 0.58 (-6) & 2.09 (-6) \\
Z  & 1.43 (-2) & 1.26 (-2) & 1.37 (-2) & 0.44 (-2) & 1.89 (-2) \\
 & & & & & \\
\hline
\end{tabular}
\end{table*}

\clearpage

\begin{figure*}
\centerline{\psfig{figure=f1.eps,width=16cm,height=16cm}} 
\end{figure*}
\vfill\eject

\begin{figure*}
\centerline{\psfig{figure=f2.eps,width=16cm,height=16cm} }
\end{figure*}

\clearpage
\eject
\vfill\eject

\begin{figure*}
\centerline{\psfig{figure=f3.eps,width=16cm,height=16cm} }
\end{figure*}

\clearpage
\eject

\begin{figure*}
\centerline{\psfig{figure=f4.eps,width=16cm,height=16cm}}
\end{figure*}

\clearpage
\eject

\begin{figure*}
\centerline{\psfig{figure=f5.eps,width=16cm,height=16cm}}
\end{figure*}

\clearpage
\eject

\begin{figure*}
\centerline{\psfig{figure=f6.eps,width=16cm,height=16cm}}
\end{figure*}

\clearpage
\eject

\begin{figure*}
\centerline{\psfig{figure=f7.eps,width=16cm,height=16cm}}
\end{figure*}

\clearpage
\eject

\begin{figure*}
\centerline{\psfig{figure=f8.eps,width=16cm,height=16cm}}
\end{figure*}

\newpage

\begin{figure*}
\centerline{\psfig{figure=f9.eps,width=16cm,height=16cm}}
\end{figure*}

\newpage

\begin{figure*}
\centerline{\psfig{figure=f10.eps,width=16cm,height=16cm}}
\end{figure*}

\newpage

\begin{figure*}
\centerline{\psfig{figure=f11.eps,width=16cm,height=16cm}}
\end{figure*}

\newpage

\begin{figure*}
\centerline{\psfig{figure=f12.eps,width=16cm,height=16cm}}
\end{figure*}

\newpage
\begin{figure*}
\centerline{\psfig{figure=f13.eps,width=16cm,height=16cm}}
\end{figure*}


\begin{thebibliography}{}

\bibitem[]{} Anders, E. \& Grevesse, N. 1989, Geochim. Cosmochim. Acta 53, 197
\bibitem[]{} Bahcall, J. N. \& Casertano, S. 1984, ApJ, 284, L35
\bibitem[]{} Blitz, L., Spergel, D. N., Teuben, P.J., Hartmann, D. \& Burton, W. B. 1999, 
ApJ, 514, 818
\bibitem[]{} Bregman, J. N. 1980, ApJ, 236, 577
\bibitem[]{} Carigi, L. 1996, Rev. Mex. Astron. Astrof., 32, 179
\bibitem[]{} Chiappini, C., Matteucci, F., Beers, T. \& Nomoto, K. 1999, ApJ, 515, 226
\bibitem[]{} Chiappini, C., Matteucci, F. \& Gratton, G. 1997, ApJ, 477, 765
\bibitem[]{} Chiappini, C. \& Maciel, W. J. 1994, A\&A, 288, 921
\bibitem[]{} Chuvenkov, V. \& Glukhov, A. 1995, ESO Astrophysics Symposium: The light element abundances , ed. P. Crane, p. 124
\bibitem[]{} Dame, T. M., Elmegreen, B. G., Cohen, R. S. \& Thaddeus, P. 1986,
ApJ, 305, 892  
\bibitem[]{} de Grijs, R. \& Peletier, R. F. 1997, A\&A, 320, L21
\bibitem[]{} Dearborn, D. S. P., Steigman, G. \& Tosi, M. 1996, ApJ, 473, 570
\bibitem[]{} Dickey, J. M. \& Lockman, F. J. 1990, ARA\&A, 28, 215  
\bibitem[]{} Edmunds, M. G.\& Greenhow, R. M. 1995, MNRAS, 272, 241
\bibitem[]{} Ferrara, A. 1993, ApJ, 407, 157
\bibitem[]{} Ferrara, A. 1996, in Unsolved Problems in the Milky Way,
eds. L. Blitz \& P. Teuben, 479
\bibitem[]{} Ferrini, F., Palla, F. \& Penco, U. 1990, A\&A, 231, 391
\bibitem[]{} Garc\'{\i}a-Burillo, S., Dahlem, M. \& Gu\'{e}lin, M. 1991,
in The Interstellar Disk--Halo Connection in Galaxies, ed. H. Bloemen, 299  
\bibitem[]{} Gerritsen, J. P. E., Icke, V 1997, A\&A, 325, 972
\bibitem[]{} Grabelsky, D. A., Cohen, R. S., Bronfman, L., Thaddeus, P. \& 
May, J. 1987, ApJ, 315, 122 
\bibitem[]{} Gratton, R., Carretta, E., Matteucci, F. \& Sneden, C. 1997 (submitted)
\bibitem[]{} Heiles, C. 1991, in The Interstellar Disk--Halo Connection
in Galaxies, ed. H. Bloemen, 433
\bibitem[]{} Kennicutt, R.C. 1989, ApJ, 344, 685
\bibitem[]{} Kennicutt, R.C. 1998, ApJ, 498, 541
\bibitem[]{} Kroupa, P., Tout, C. A. \& Gilmore, G. 1993, MNRAS, 262, 545
\bibitem[]{} Larson, R. B. 1998, MNRAS, 301, 569
\bibitem[]{} Larson, R. B. 1976, MNRAS, 176, 31
\bibitem[]{} Larson, R. B. 1981, MNRAS, 194, 809
\bibitem[]{} McGaugh, S. \& Blok, E. 1996 (preprint) - astro-ph/9612019
\bibitem[]{} Maciel, W. J.\& Quireza, C. 1999, A\&A, 345, 629
\bibitem[]{} Martinelli, A. \& Matteucci, F. 1999, A\&A (in press) 
\bibitem[]{} Matteucci, F. 1996, Fund. Cosm. Physics, 17, 283
\bibitem[]{} Matteucci, F. \& Brocato, E. 1990, ApJ, 365, 539
\bibitem[]{} Matteucci, F., Ferrini, F., Pardi, C. \& Penco, U. 1990, in Chemical and dynamical evolution of galaxies, eds. F. Ferrini, J. Franco and F. Matteucci, ETS Editrice, PISA, p. 586
\bibitem[]{} Matteucci, F. \& Fran\c cois, P. 1989, MNRAS, 239, 885
\bibitem[]{} Matteucci, F. \& Tornamb\'e, A. 1985, A\&A, 142, 13
\bibitem[]{} Miller, G. E. \& Scalo, J. M. 1979, ApJS, 41,453
\bibitem[]{} Mooney, T. J. \& Solomon, P. M. 1988, ApJ, 334, L51
\bibitem[]{} Myers, P. C., Dame, T. M., Thaddeus, P., Cohen, R. S., Silverberg, L. F., Dwek, E. \& Hauser, M. G. 1986, ApJ, 301, 398
\bibitem[]{} Nordlund, A. A. \& Padoan, P. 1997 (in preparation)
\bibitem[]{} Padoan, P., Nordlund, A. A. \& Jones, B. J. T. 1997, MNRAS, 288, 145 (PNJ)
\bibitem[]{} Pagel, B. E. J. 1997, Nucleosynthesis and chemical evolution of galaxies, Cambridge University Press
\bibitem[]{} Pagel, B. E. J. \& Tautvaisiene, G. 1995, MNRAS, 276,505
\bibitem[]{} Pagel, B. E. J., Simonson, E. A., Terlevich, R. J. \& Edmunds, M.G. 1992, MNRAS, 255, 325
\bibitem[]{} Pagel, B. E. J. \& Patchett, B. E. 1975, MNRAS, 172, 13
\bibitem[]{} Prantzos, N. \& Aubert, O. 1995, A\&A, 302, 69
\bibitem[]{} Rana, N. 1991, ARA\&A, 29,129
\bibitem[]{} Renzini, A. \& Voli, M. 1981, A\&A, 94, 175
\bibitem[]{} Rocha-Pinto, H. J. \& Maciel, W. J. 1996, MNRAS, 279,447
\bibitem[]{} Roche, N., Ratnatunga, K., Griffiths, R. E., Im, M. \& Naim, A. 1998, MNRAS 293, 157
\bibitem[]{} Salpeter, E. E. 1955, ApJ, 121, 161
\bibitem[]{} Sanders, D. B., Solomon, P. M. \& Scoville, N. Z. 1984, ApJ, 276, 182
\bibitem[]{} Scalo, J. M. 1998, 38th Hertmonceux Conference, eds. G. Gilmore and D. Howell, ASP Conf. Series, Vol. 142, p.201
\bibitem[]{} Scalo, J. M., Vazquez-Semadeni, E., Chappell, D. \& Passot, T. 1998, ApJ, 504, 835
\bibitem[]{} Scalo, J. M. 1986, Fund. Cosm. Phys., 11, 1
\bibitem[]{} Scully, S., Cass\'e, M., Olive, K. A. \& Vangioni-Flam, E. 1996, ApJ, 462, 960
\bibitem[]{} Silk, J. 1995, ApJ, 438, L41
\bibitem[]{} Solomon, P.M., Sanders, D.B., \& Scoville, N.Z. 
1979, ApJ, 232, L98
\bibitem[]{} Terlevich, R. J. \& Melnick, J. 1983, E.S.O. preprint n. 264
\bibitem[]{} Tinsley, B. M. 1980, Fundamentals of Cosmic Physics, 5, 287
\bibitem[]{} Thielemann, F. K., Nomoto, K. \& Hashimoto, M. 1993 in Origin and Evolution of the Elements, eds Prantzos, Vangioni-Flam, Cass\'e, p. 297
\bibitem[]{} Tosi, M. 1996 in ASP Conf. Proc. 98, From Stars to Galaxies, eds. C. Leitherer, U. Fritze-von Alvensleben and J. Huchra (San Francisco: ASP), p. 299
\bibitem[]{} Wielen, R. \& Wilson, T. L. 1997, A\&A, 326, 139
\bibitem[]{} Woosley, S. E. \& Weaver, T. A. 1995, ApJS, 101, 181

\end{thebibliography}
\end{document}